\documentclass[aps,prl,twocolumn,superscriptaddress]{revtex4-1}

\usepackage{amssymb}
\usepackage{natbib}
\usepackage{multirow}
\usepackage{bm}
\usepackage{amsmath}
\usepackage{graphicx}
\usepackage{epstopdf}
\usepackage{subfigure}
\usepackage{natbib}
\usepackage{epsfig}
\usepackage{xcolor}
\usepackage{amsfonts}
\usepackage{mathrsfs}
\usepackage{braket}
\usepackage{tabularx}
\usepackage[toc,page,title,titletoc,header]{appendix}

\usepackage{dsfont,amsthm,amsbsy}

\usepackage{lineno}
\begin{document}

\title{Mapping twist-tuned multiband topology in bilayer WSe$_2$}

\author{Benjamin A. Foutty}
\affiliation{Geballe Laboratory for Advanced Materials, Stanford, CA 94305, USA}
\affiliation{Department of Physics, Stanford University, Stanford, CA 94305, USA}

\author{Carlos R. Kometter}
\affiliation{Geballe Laboratory for Advanced Materials, Stanford, CA 94305, USA}
\affiliation{Department of Physics, Stanford University, Stanford, CA 94305, USA}

\author{Trithep Devakul}
\affiliation{Department of Physics, Massachusetts Institute of Technology, Cambridge, Massachusetts 02139, USA}

\author{Aidan P. Reddy}
\affiliation{Department of Physics, Massachusetts Institute of Technology, Cambridge, Massachusetts 02139, USA}

\author{Kenji Watanabe}
\affiliation{Research Center for Electronic and Optical Materials, National Institute for Materials Science, 1-1 Namiki, Tsukuba 305-0044, Japan}

\author{Takashi Taniguchi}
\affiliation{Research Center for Materials Nanoarchitectonics, National Institute for Materials Science, 1-1 Namiki, Tsukuba 305-0044, Japan}

\author{Liang Fu}
\affiliation{Department of Physics, Massachusetts Institute of Technology, Cambridge, Massachusetts 02139, USA}

\author{Benjamin E. Feldman}
\email{bef@stanford.edu}
\affiliation{Geballe Laboratory for Advanced Materials, Stanford, CA 94305, USA}
\affiliation{Department of Physics, Stanford University, Stanford, CA 94305, USA}
\affiliation{Stanford Institute for Materials and Energy Sciences, SLAC National Accelerator Laboratory, Menlo Park, CA 94025, USA}


\begin{abstract}
Semiconductor moir\'e superlattices have been shown to host a wide array of interaction-driven ground states. However, twisted homobilayers have been difficult to study in the limit of large moir\'e wavelength, where interactions are most dominant. Here, we conduct local electronic compressibility measurements of twisted bilayer WSe$_2$ (tWSe$_2$) at small twist angles. We demonstrate multiple topological bands which host a series of Chern insulators at zero magnetic field near a ‘magic angle’ around $1.23^\circ$. Using a locally applied electric field, we induce a topological quantum phase transition at one hole per moir\'e unit cell. Our work establishes the topological phase diagram of a generalized Kane-Mele-Hubbard model in tWSe$_2$, demonstrating a tunable platform for strongly correlated topological phases.

\end{abstract}

\maketitle

\section{Introduction}

The richness of interacting condensed matter systems motivates the development of experimental platforms with tunable Hamiltonians, which can admit systematic study over a wide range of phase space \cite{bloch_quantum_2012,kennes_moire_2021}. Recently, moir\'e heterostructures have emerged as a venue to engineer flat electronic bands and realize emergent correlated states. In twisted graphene structures, interacting phases generally only form near certain `magic' angles \cite{andrei_graphene_2020,balents_superconductivity_2020,cao_correlated_2018,cao_unconventional_2018}. In contrast, moir\'e transition metal dichalcogenide (TMD) heterostructures exhibit strong correlations across a wide range of twist angles (and corresponding moir\'e wavelengths) \cite{mak_semiconductor_2022}. Because the underlying electronic structure can often be captured by extended Hubbard models, these systems are promising for quantum simulation with effective parameters tunable by moir\'e wavelength, choice of TMD, and applied electromagnetic fields \cite{kennes_moire_2021,wu_hubbard_2018,zhang_moire_2020,tang_simulation_2020}. In practice, the ability to meaningfully adjust parameters has been limited by the experimental challenges of poor electrical contact and local disorder \cite{mak_semiconductor_2022,lau_reproducibility_2022}. As a result, study of correlated states in these systems to date has focused mostly on heterobilayers, where lattice mismatch provides a nearly constant moir\'e wavelength \cite{tang_simulation_2020,xu_correlated_2020,regan_mott_2020,li_continuous_2021,li_quantum_2021,li_charge-order-enhanced_2021}, or on homobilayers with large interlayer twist, where angular variations have a muted effect \cite{wang_correlated_2020,ghiotto_quantum_2021,xu_tunable_2022}. While this has resulted in the discovery of many exotic ground states, the ability to vary parameters while maintaining strong interactions relative to electronic bandwidth in a single system has remained limited.

Of particular interest is the ability to engineer tunable topological flat-band structures, in which interactions can lead to nontrivial insulating states. In graphene-based moir\'e superlattices, quantum anomalous Hall (QAH) insulators have been observed, but this typically requires fine-tuning of external parameters (in addition to twist angle) \cite{serlin_intrinsic_2020,chen_tunable_2020,pierce_unconventional_2021,polshyn_electrical_2020,stepanov_competing_2021}, and the topological bands do not allow a local tight-binding description \cite{andrei_graphene_2020,mak_semiconductor_2022,kane_quantum_2005}. 
Theoretical calculations have predicted that certain TMD moir\'e homobilayers realize a generalized version of the Kane-Mele model in which the lowest energy pair of bands have nonzero Chern number, and that these bands could support QAH states at odd filling thanks to interactions \cite{wu_topological_2019}. However, there has been disagreement in the literature about the exact parameters for which topological bands are expected \cite{devakul_magic_2021,pan_band_2020,mai_14_2023}. Although there has been no experimental evidence of topology in TMD homobilayers, recent experiments in AB-stacked MoTe$_2$/WSe$_2$ heterobilayers were able to mimic similar physics using a large out of plane electric field to hybridize the electronic states on the two layers, and realize QAH and quantum spin hall insulating states at moir\'e filling factors $\nu = -1$ and $\nu = -2$, respectively \cite{li_quantum_2021,tschirhart_intrinsic_2023,zhao_realization_2022}. The precise symmetry-breaking that underlies these insulating phases is a subject of ongoing experimental and theoretical studies \cite{tao_valley-coherent_2022,zhang_spin-textured_2021,xie_valley-polarized_2022,pan_topological_2022,devakul_quantum_2022,rademaker_spin-orbit_2022,dong_excitonic_2023,mai_interaction-driven_2023,mai_14_2023}.

\begin{figure*}[t!]
    \renewcommand{\thefigure}{\arabic{figure}}
    \centering
    \includegraphics[scale =1.0]{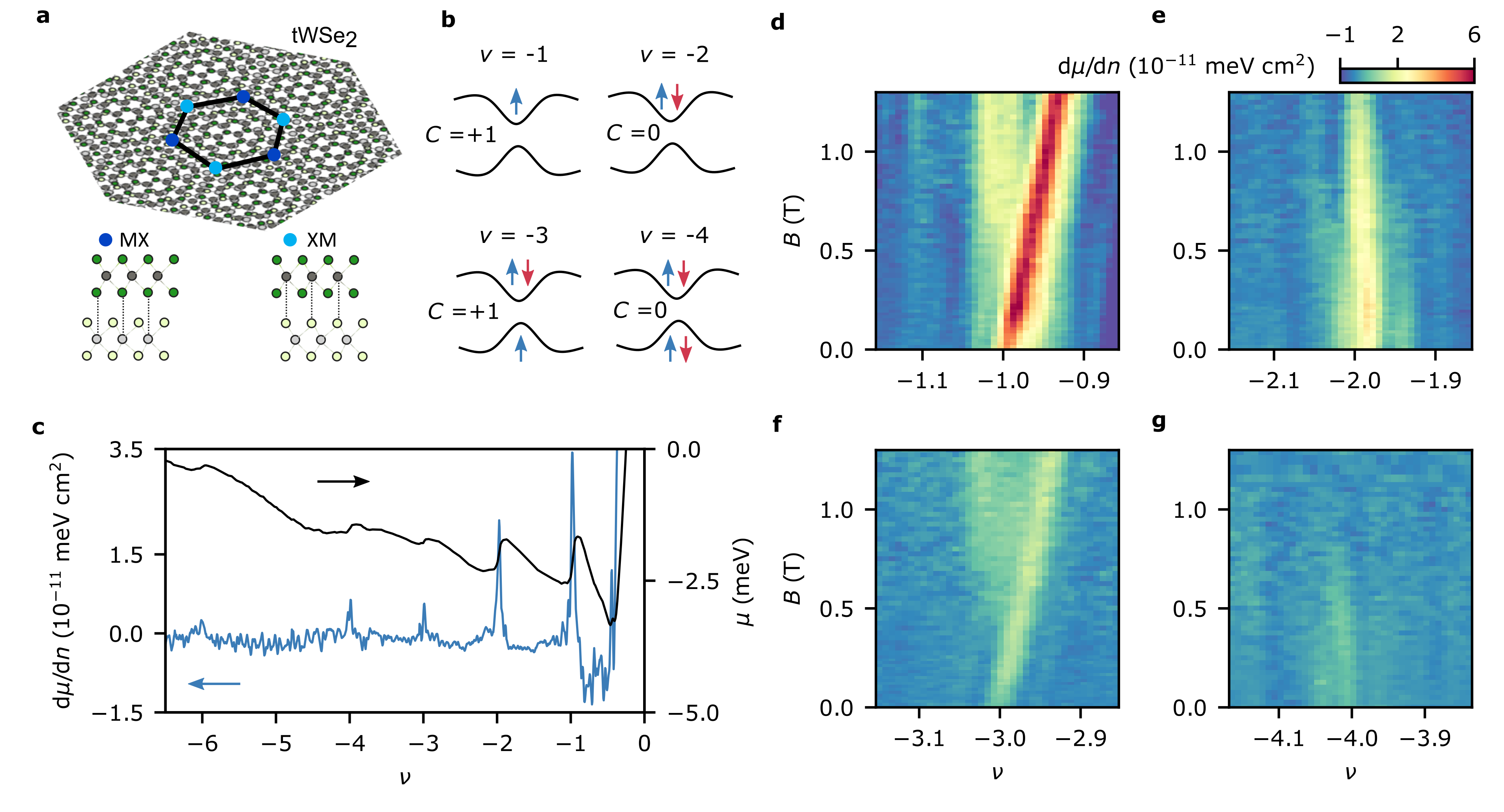}
    \caption{\textbf{Multiple topological bands and ground states in tWSe$_2$.} \textbf{a}, Schematic showing the moir\'e pattern in twisted WSe$_2$ (tWSe$_2$), in which MX and XM stacking sites (bottom illustrations) form a honeycomb lattice. \textbf{b}, Cartoon of lowest energy moir\'e bands and their occupations at the first four integer hole fillings, with Chern numbers $C$ of the corresponding ground states labeled. \textbf{c}, Measurement of inverse electronic compressibility $\mathrm{d}\mu/\mathrm{d}n$ and chemical potential $\mu(n)$ of tWSe$_2$ as function of holes per moir\'e unit cell $\nu$ in a location with twist angle $\theta = 1.23^\circ$. \textbf{d-g}, $\mathrm{d}\mu/\mathrm{d}n$ at low magnetic fields $B$ around each of the first four integer moir\'e fillings $\nu$. The gaps at $\nu = -1$ and $-3$ have $C=+1$, while $\nu = -2$ and $-4$ have $C=0$.}
    \label{fig:Fig1}
\end{figure*}

Here we use scanning single electron transistor (SET) microscopy to investigate twisted bilayer WSe$_2$ (tWSe$_2$) at much lower relative twist angle $\theta$ than previously studied. By measuring at different twist angles in far-separated locations, we map how the system depends on local moir\'e wavelength. In addition to widespread correlated insulating states, we find `magic-angle' topology in the vicinity of $\theta \approx 1.23^\circ$. Around that twist angle, we observe a series of QAH insulators, indicating multiple moir\'e bands with nonzero Chern numbers. Using an applied displacement field, we drive a topological phase transition of the $\nu = -1$ insulator between a QAH state and a topologically trivial layer-polarized state. Together, our results represent systematic quantum simulation of generalized Kane-Mele-Hubbard physics on a tunable real-space lattice.

\section{Series of Chern insulators in tWSe$_2$}

We study a twisted WSe$_2$ bilayer with a small relative angle $1.1^\circ < \theta < 1.6^\circ$. At these twist angles, the lowest two moir\'e valence bands are well-described by a tight-binding model on a honeycomb lattice consisting of states localized at the XM and MX sites, shown schematically in Fig. \ref{fig:Fig1}a \cite{devakul_magic_2021}. We first focus on an area of the sample with $\theta = 1.23^\circ$. Figure \ref{fig:Fig1}c shows the chemical potential $\mu$ and the inverse electronic compressibility d$\mu/$d$n$ as a function of moir\'e filling factor $\nu$, where $\nu = -1$ signifies one hole per moir\'e unit cell. Gapped states occur at $\nu = -1,-2,-3,-4$, and $-6$, superimposed on an overall downslope in $\mu$, which reflects long-range electronic interactions near the valence band edge \cite{eisenstein_negative_1992, foutty_tunable_2023}. 
\begin{figure*}[t!]
    \renewcommand{\thefigure}{\arabic{figure}}
    \centering
    \includegraphics[scale=1.0]{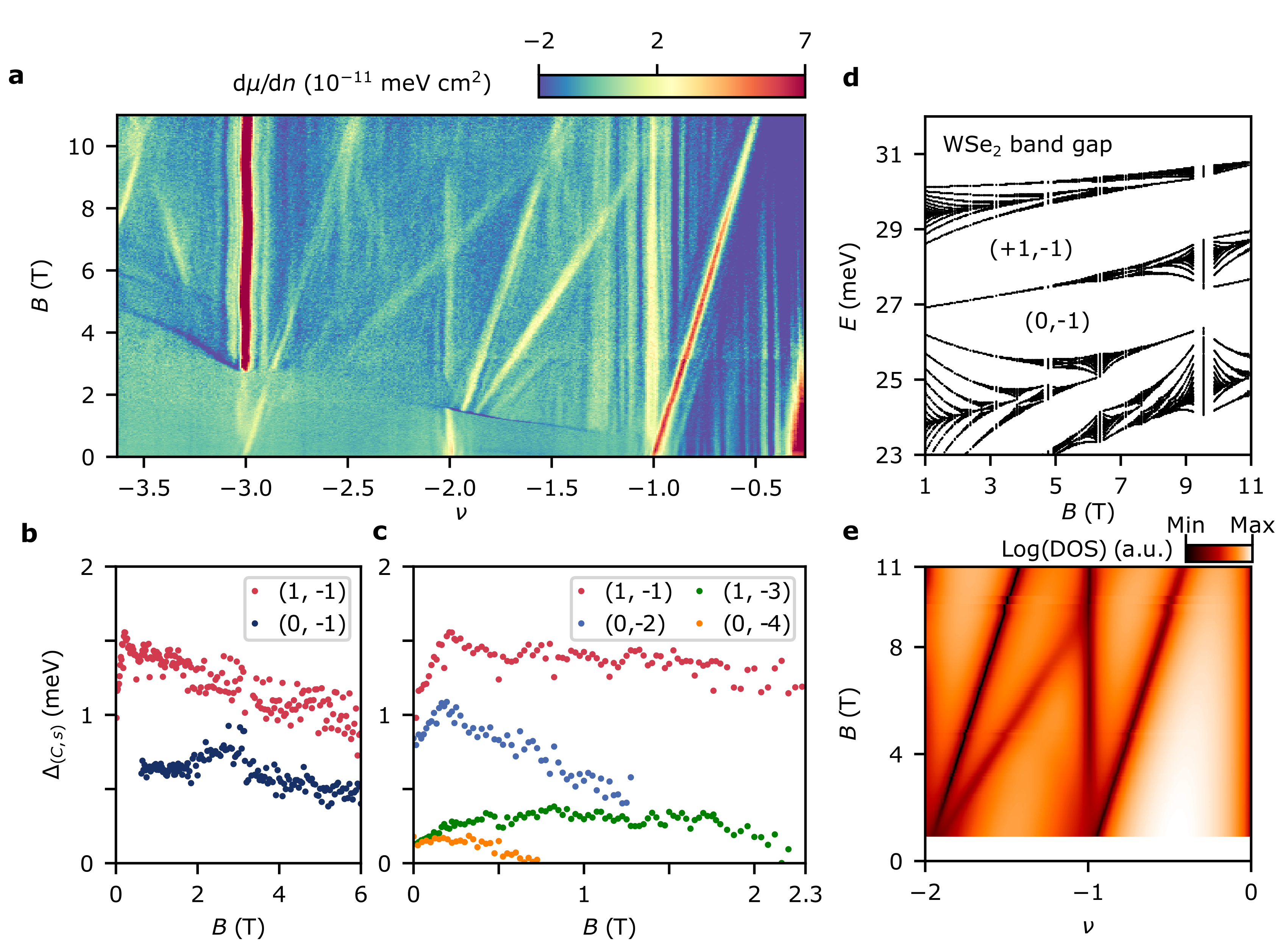}
    \caption{\textbf{Magnetic field dependence at $\theta = 1.23^\circ$.} \textbf{a} Magnetic field dependence of d$\mu$/d$n$ up to $B = 11$ T. 
    \textbf{b}, Thermodynamic gaps $\Delta_{(C,s)}$ for states with intercept $s$ and slope $C$ as identified by the Diophantine equation, as a function of $B$, for the pair of competing states with intercept $s = -1$. \textbf{c}, $\Delta_{(C,s)}$ as a function of $B$, for the insulating gaps that persist to zero magnetic field at $s = -1,-2,-3$, and $-4$. \textbf{d}, Theoretically calculated single-particle Hofstadter spectrum for a single-spin species as a function of $B$ near a topological band inversion, with single-particle energy $E$ measured relative to the WSe$_2$ band edge. The prominent $C = +1$ and $C =0$ gaps with $s = -1$ are labelled. \textbf{e}, Wannier plot of the density of states of the Hofstadter spectrum in \textbf{d} as a function of $\nu$ and $B$, with $0.5$ meV Gaussian broadening.}
    \label{fig:Fig2}
\end{figure*}

To determine the nature of the insulating states, we probe the behavior near integer fillings $|\nu|\le4$ at low magnetic field $B$ and temperature $T  = 330$ mK (Fig. \ref{fig:Fig1}d-g). The slopes of the gaps encode their topological character (and Hall conductance) via the Streda formula \cite{streda_theory_1982}. At $\nu = -1$ and $-3$, the gaps have slopes corresponding to Chern number $C = +1$ all the way down to zero magnetic field, indicating that both are QAH states (Sec. 1 of \cite{Supp}). These QAH states likely arise from half-filling and polarizing underlying Chern moir\'e bands \cite{wu_topological_2019,devakul_magic_2021} (see Sec. 2 of \cite{Supp} for further discussion). At the even integers $\nu = -2$ and $-4$, we measure gaps with total Chern number $C=0$. Based on the observed nontrivial topology at odd integer fillings, these states could be quantum spin Hall insulators with either two or four copies of fully filled $|C| = 1$ bands (Fig. \ref{fig:Fig1}b), but numerics suggest that an in-plane antiferromagnetic state is also competitive \cite{wu_topological_2019,qiu_interaction-driven_2023}. We measure broadly similar behavior at $T = 1.6$ K (Fig. S1) and across a wide spatial region (roughly 600 nm $\times$ 1 $\mu$m) in our sample with uniform twist angle between $1.22^\circ$ and $1.23^\circ$ (Fig. S2).

The observation of multiple topological bands that are not related by degenerate quantum degrees of freedom appears to be unique among moir\'e systems. Our experiment is most consistent with the pair of lowest energy moir\'e bands having identical Chern number in a single spin/valley sector (Sec. 2-3 of \cite{Supp}). This is beyond any two-band tight binding model (which must have zero total Chern number), highlighting that the topological structure in tWSe$_2$ extends beyond the Kane-Mele model \cite{wu_topological_2019,mai_14_2023,qiu_interaction-driven_2023}.

In Fig. \ref{fig:Fig2}a, we present the full magnetic field dependence at this twist angle up to $B = 11$ T. A feature of both Chern insulating states at $\nu = -1$ and $\nu = -3$ is the coexistence of topologically trivial $C=0$ states emerging at relatively low magnetic fields $B \approx 0.6$ T, where no other Landau levels (Hofstadter states) are seen. The thermodynamic charge gap of the trivial state is smaller than the topological state, and both gaps shrink in moderate magnetic fields (Fig. \ref{fig:Fig2}b). This behavior is consistent with theoretical expectations for Chern bands close to a topological band inversion between the first and second moir\'e bands (Fig. \ref{fig:Fig2}d-e, Sec. 3 of \cite{Supp}), for which a Landau level is pinned to the top of the second moir\'e band, leading to two dominant Hofstadter gaps originating from $\nu=-1$.

At moderate magnetic fields, multiple additional Hofstadter states emerge. Each of these can be described by the Diophantine equation $\nu = C(\Phi/\Phi_0) + s$, where $C$ is the Chern number of the gap, $\Phi$ is the magnetic flux per moir\'e unit ceell, $\Phi_0 = h/e$ is the flux quantum, and $s$ is the intercept at zero magnetic field. In addition to the Hofstadter states, we observe phase transitions as a function of magnetic field (and density), signified by sharp negative compressibility and changes in the pattern and/or strength of gaps such as that near $\nu = -3$ at $B = 2.6$ T. This likely reflects changing energetics between distinct moir\'e bands in a magnetic field \cite{cao_correlated_2018,kometter_hofstadter_2022}.

The magnitudes of the dominant gaps emanating from integer fillings exhibit nontrivial dependence at low magnetic fields (Fig. \ref{fig:Fig2}c). All of these gaps are non-monotonic below $B = 1$ T. This is particularly stark at $\nu = -1$ and $-2$, where the gaps sharply increase up to $B \approx 0.25$ T, before decreasing beyond that field. One explanation for this gap dependence might be changes in the spin ordering of the ground states at very low magnetic fields. However, given the topology of the $\nu = -1$ gap and the theoretically predicted spin/valley polarization of the QAH state, this is unlikely. Another possibility is a change in the nature of the lowest energy excitations, such as their spin configuration, as a field is applied. We note that the non-monotonic dependence of the gaps at $\nu = -1$ and $\nu = -2$ persists to angles where all states are topologically trivial (Fig. S4).

\section{Twist angle dependence}

\begin{figure}[h]
    \renewcommand{\thefigure}{\arabic{figure}}
    \centering
    \includegraphics[scale=1]{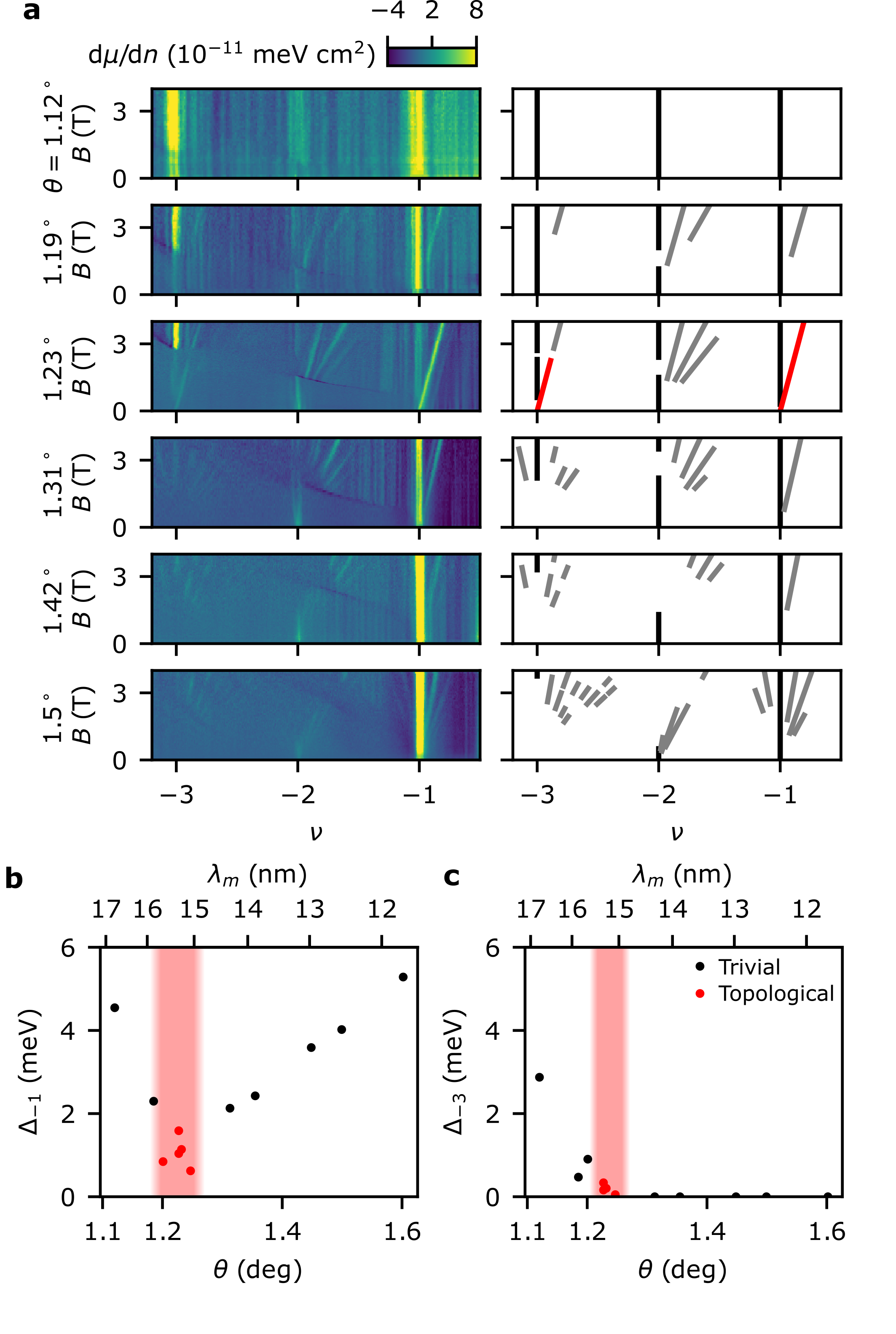}
    \caption{\textbf{Twist angle dependence of correlated states.} \textbf{a}, Left: $\mathrm{d}\mu/\mathrm{d}n$ as a function of $\nu$ and $B$ at a selection of twist angles. From top to bottom, $\theta = 1.12^\circ$, $1.19^\circ$, $1.23^\circ$, $1.31^\circ$, $1.42^\circ$, and $1.50^\circ$. Right: corresponding Wannier plot for each angle. Black lines show states with $C=0$, gray lines show (Hofstadter) states with $C \neq 0$ that do not persist to $B = 0$, red lines show states with $C\neq 0$ that persist to $B = 0$. \textbf{b-c}, Thermodynamic gaps $\Delta_{\nu}$ at $B=0$ for fillings $\nu = -1$ (\textbf{b}) and $\nu = -3$ (\textbf{c}) as a function of twist angle (bottom axis) and moir\'e wavelength $\lambda_m$ (top axis). Red dots and shading indicate where the gap is topological.}
    \label{fig:Fig3}
\end{figure}

Although the twist angle is homogeneous within domains that are typically much larger than the $\sim 100$ nm lengthscale of our probe, we find that different regions of our device have different local twist angles (Fig. S2). Measurements at other locations with local angles ranging from $\theta\sim 1.1-1.6^{\circ}$ enable us to systematically investigate how these states depend on moir\'e wavelength. We present d$\mu$/d$n$ as a function of $\nu$ up to $B = 4$ T at a selection of twist angles in Fig. \ref{fig:Fig3}a, with the observed incompressible states schematically depicted in the adjacent Wannier diagrams. Outside of a small range of angles ($\theta = 1.2^\circ - 1.25^\circ$) we do not observe any zero-field Chern insulators. At $\theta = 1.20^\circ$, we find that a QAH state is favored at $\nu = -1$, but not $\nu = -3$, indicating a slight difference in the twist angle dependence of these two states (Fig. S5). Qualitatively, below this range, the integer gaps become much stronger, and fewer Hofstadter states are observed. Additionally there is a reversal of the relative gap strengths: at low twist angle, $\Delta_{-3}$ becomes larger than the $\Delta_{-2}$, likely owing to details of the interaction strength relative to the single-particle gap between moir\'e bands. The overall level of disorder also consistently appears worse, which may reflect an enhanced susceptibility to local atomic reconstruction at low angles. For $\theta > 1.3^\circ$, no zero-field gaps are observed at $|\nu| \ge 3$, and the $\nu = -2$ gap vanishes at lower magnetic fields with increasing angle. Additionally, we observe a larger number of (weaker) Hofstadter states tracing back to all integers.

These data, along with more measurements at intermediate twists (Figs. S3,S6), allow us to quantify gap size dependence on twist angle at zero magnetic field, and to correlate it with the nature of the ground states that we observe. At $\nu = -1$, topological gaps appear at an overall minimum of a non-monotonic dependence of $\Delta_{-1}$ as a function of twist angle (Fig. \ref{fig:Fig3}b). This is consistent with a pair of twist-tuned band inversion, where we would expect the thermodynamic gaps to close. Additionally, the non-monotonic angle dependence of the gap outside the topological regime suggests that the trivial insulating phases on either side have different spin/valley ordering, which is consistent with our Hartree-Fock calculations (Sec. 2 of \cite{Supp}).

In contrast, the trend at $\nu = -3$ is generally monotonic as a function of angle, and the gap only becomes topological just as it closes (Fig. \ref{fig:Fig3}c). At higher twist angles, interactions are likely not strong enough to induce an insulating state at this filling, so we realize a (trivial) insulator to QAH insulator to metal transition as a function of increasing twist angle at $\nu = -3$.

 \begin{figure*}[t!]
    \renewcommand{\thefigure}{\arabic{figure}}
    \centering
    \includegraphics[scale=1.0]{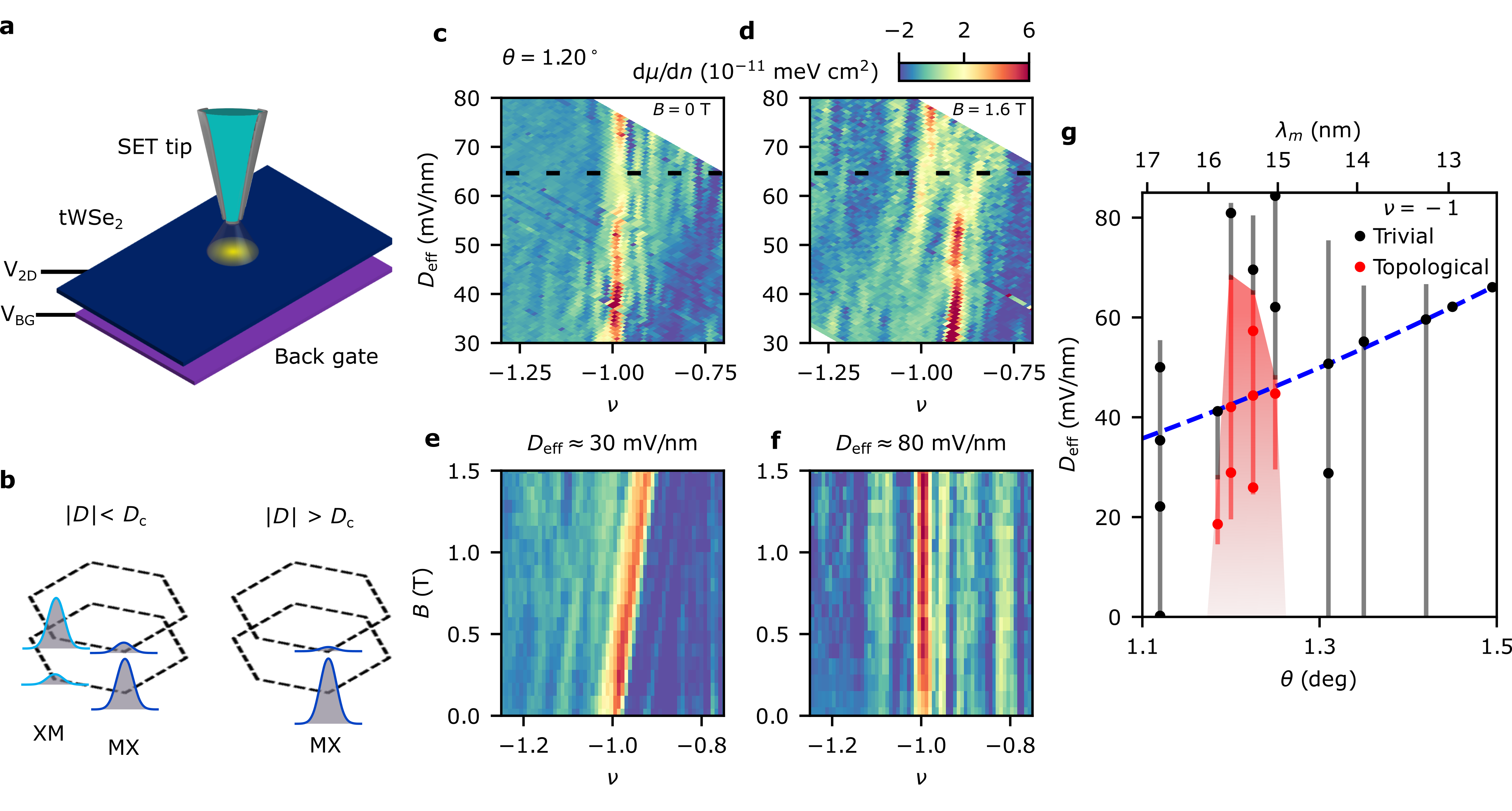}
    \caption{\textbf{Electric field tuning and topological phase diagram at $\nu = -1$.} \textbf{a}, Schematic of the single-electron transistor (SET) tip above a tWSe$_2$ sample with a global back gate. The effective top gate is set by the voltage applied to the sample $V_{2D}$ relative to the grounded tip, while the effective back gate voltage is given by the difference between the back gate voltage $V_{BG}$ and $V_{2D}$. \textbf{b}, Cartoon of spatial and layer-resolved density within a moir\'e unit cell. At low $|D|$, states are evenly spread across the XM and MX sites, forming a honeycomb lattice. At high $D$, holes preferentially occupy a single layer at one of these two sites, chosen by the sign of $D$. \textbf{c-d}, d$\mu$/d$n$ as a function of $D_{\rm{eff}}$ and $\nu$ at $B = 0$ T (\textbf{c}) and $B = 1.6$ T (\textbf{d}) in a location with $\theta = 1.2^\circ$. The black dashed line marks the critical field $D_c$ of a topological transition. \textbf{e}, d$\mu$/d$n$ as a function of $\nu$ and $B$ around $\nu = -1$ at $V_{2D}=3.2$ V in which the tip induces a reduced effective displacement field $D_{\rm{eff},\nu=-1} \approx 30$ mV/nm. The resulting gap is topological with Chern number $C=+1$. \textbf{f}, d$\mu$/d$n$ as a function of $\nu$ and $B$ around $\nu = -1$ at $V_{2D}=-2$ V in which the tip induces a higher effective displacement field $D_{\rm{eff},\nu=-1} \approx 80$ mV/nm. The resulting gap is trivial with Chern number $C=0$. \textbf{g}, Experimental phase diagram of the topological character of the $\nu = -1$ gap as a function of twist angle (top axis) or moir\'e wavelength (bottom axis) and effective displacement field. Red indicates topological ($C = +1$) while black indicates trivial. Dots represent locations/conditions where we have conducted field sweeps (as in \textbf{e-f}), while lines indicate measurements as a function of $n$ and $D_{\rm{eff}}$ at fixed $B$ (as in \textbf{c-d}). The blue dashed line indicates the condition where the tip does not dope the sample.} 
    \label{fig:Fig4}
\end{figure*}
\section{Topological phase transition at $\nu = -1$}

Theoretical predictions assert that the topological phase at $\nu = -1$ is highly sensitive to an applied electric field, which favors polarization onto one of the layers and breaks the honeycomb sublattice symmetry of the emergent generalized Kane-Mele bands (Fig. \ref{fig:Fig4}b) \cite{devakul_magic_2021}. Though the scanning SET tip is generally noninvasive, if the d.c. sample voltage $V_{2D}$ is adjusted, the tip can locally dope the region underneath it. We find that within certain bounds of $V_{2D}$, the tip acts as an effective gate that tunes the local displacement field $D_{\rm{eff}}$ applied to the sample (Fig. \ref{fig:Fig4}a, Methods). Notably, we observe a transition at $\nu = -1$ as a function of $D_{\rm{eff}}$ at certain twist angles.

We focus on a location with $\theta = 1.20^\circ$. At $B = 0$, the gap at $\nu = -1$ first decreases before re-emerging and strengthening as $D_{\rm{eff}}$ is increased (Fig. \ref{fig:Fig4}c). At $B = 1.6$ T, states with different $C$ occur at different density. At lower $D_{\rm{eff}}$, the $(C,s) = (+1, -1)$ state is dominant, while the $(0,-1)$ state strengthens at high $D_{\rm{eff}}$ (Fig. \ref{fig:Fig4}d). At both magnetic fields, the change in behavior occurs at a similar critical displacement field around $D_c = 65$ mV/nm (Fig. \ref{fig:Fig4}c-d, black dashed line). This suggests a phase transition from a topological state below $D_c$ to a trivial $\nu=-1$ gap at higher displacement fields. The observations at both magnetic fields are consistent with the thermodynamic gap closing and reopening across $D_c$, as expected from a topological quantum phase transition \cite{hasan_colloquium_2010,li_quantum_2021}.

To confirm this interpretation, we measure d$\mu$/d$n$ as a function of magnetic field at different effective displacement fields. In Fig. \ref{fig:Fig4}e, we present a measurement at the same location with a small displacement field $D_{\rm{eff}} \approx 30$ mV/nm. The resulting $\nu = -1$ state is topological, with the $C = +1$ gap continuing down to $B=0$. However, when the displacement field is increased to $D_{\rm{eff}} \approx 80$ mV/nm, a field sweep in the same location shows a topologically trivial gap (Fig. \ref{fig:Fig4}f). 

By repeating similar measurements at other locations, we build up a phase diagram of the topological behavior at $\nu = -1$ as a function of twist angle and applied displacement field (Fig. \ref{fig:Fig4}g). In this plot, vertical lines signify measurements as a function of $D_{\rm{eff}}$, while points signify field sweeps to confirm the topological character of each state. In addition to being able to tune the topological gaps to trivial at $\theta = 1.20^\circ - 1.25^\circ$, we are also able to induce topological states where trivial states had been favored (at $\theta = 1.19^\circ$) by decreasing the applied displacement field (Sec. 4 of \cite{Supp}). The range of twist angles over which we observe topological behavior is quite narrow, likely because of other competitive states at $\nu = -1$.

We perform Hartree-Fock calculations for the continuum model (Sec. 2 of \cite{Supp}) which indicate a rich landscape of correlated ground states at integer $\nu$. Within a certain range of twist angles at $\nu=-1$, a spin-valley polarized Chern insulator is expected to be favored, with transitions to a topologically trivial layer-polarized state at sufficiently high $D$ or low $\theta$. At higher twist angle, our calculations predict that the Chern state will give way to an intervalley-coherent trivial state. Additionally, we find a quantum spin Hall insulator at $\nu=-2$, and a Chern insulator at $\nu=-3$ with the same sign as at $\nu=-1$, as observed in experiment.

\section{Discussion and outlook}

In conclusion, we have established twisted WSe$_2$ as a platform to realize multiple topologically nontrivial bands that support symmetry-broken QAH gaps at half-filling. Comparing data across a range of moir\'e wavelengths from approximately 12 to 17 nm, we reveal widely varying correlated ground states of the honeycomb lattice in twisted WSe$_2$, both in magnitudes of the gaps and their topological character. This highlights the importance of moir\'e wavelength as a tuning parameter, which is only accessible in low-twist lattice-matched systems. Our work motivates further studies to confirm the specific spin/valley symmetry breaking of the odd-integer gaps, as well as transport studies to confirm whether quantum spin Hall insulating states are realized at even integers, including a potential double quantum spin Hall insulator at $\nu = -4$ \cite{mai_14_2023}. More generally, this work opens a new platform for studying tunable topological bands in the limit of larger moir\'e wavelengths where interacting phases such as fractional Chern insulators may be more readily stabilized \cite{crepel_anomalous_2022,morales-duran_pressure--enhanced_2023}.  \\

\noindent
\textit{Note added}: While under review, we became aware of related work observing QAH states at $\nu = -1$ as well as fractional QAH states in twisted bilayer MoTe$_2$ \cite{cai_signatures_2023,zeng_thermodynamic_2023,park_observation_2023,xu_observation_2023}.


\section{Data and code availability}
Data from the main text and Supplementary materials and code used to for data analysis is presented in a Zenodo repository \cite{Zenodo}.

\section{Acknowledgements}

Experimental work was primarily supported by the Department of Energy, Office of Basic Energy Sciences, award number DE-SC0023109. B.E.F. acknowledges an Alfred P. Sloan Foundation Fellowship and a Cottrell Scholar Award. The work at Massachusetts Institute of Technology was supported by the Air Force Office of Scientific Research (AFOSR) under award FA9550-22-1-0432. K.W. and T.T. acknowledge support from the JSPS KAKENHI (Grant Numbers 20H00354 and 23H02052) and World Premier International Research Center Initiative (WPI), MEXT, Japan. B.A.F. acknowledges a Stanford Graduate Fellowship. Part of this work was performed at the Stanford Nano Shared Facilities (SNSF), supported by the National Science Foundation under award ECCS-2026822.

\section{Author contribution}
B.A.F. and C.R.K. conducted the scanning SET measurements. B.A.F. and B.E.F. designed the experiment. T.D., A.P.R., and L.F. conducted theoretical calculations. B.A.F. fabricated the sample. K.W. and T.T. provided the hBN crystals. B.E.F. and L.F. supervised the project. All authors participated in analysis of the data and writing of the manuscript. 

\section{Competing interests}
The authors declare no competing interest. 

\section{List of supplementary materials}
\begin{itemize}
\item{Materials and Methods}
\item{Supplementary Text (Secs. 1-7)}
\item{Supplementary Figures (S1-S16)}
\item{References (54-75)}
\end{itemize}

\end{document}


\title{Supplementary Material for:\\Mapping twist-tuned multiband topology in bilayer WSe$_2$}

\author{Benjamin A. Foutty}
\affiliation{Geballe Laboratory for Advanced Materials, Stanford, CA 94305, USA}
\affiliation{Department of Physics, Stanford University, Stanford, CA 94305, USA}

\author{Carlos R. Kometter}
\affiliation{Geballe Laboratory for Advanced Materials, Stanford, CA 94305, USA}
\affiliation{Department of Physics, Stanford University, Stanford, CA 94305, USA}

\author{Trithep Devakul}
\affiliation{Department of Physics, Massachusetts Institute of Technology, Cambridge, Massachusetts 02139, USA}

\author{Aidan P. Reddy}
\affiliation{Department of Physics, Massachusetts Institute of Technology, Cambridge, Massachusetts 02139, USA}

\author{Kenji Watanabe}
\affiliation{Research Center for Electronic and Optical Materials, National Institute for Materials Science, 1-1 Namiki, Tsukuba 305-0044, Japan}

\author{Takashi Taniguchi}
\affiliation{Research Center for Materials Nanoarchitectonics, National Institute for Materials Science, 1-1 Namiki, Tsukuba 305-0044, Japan}

\author{Liang Fu}
\affiliation{Department of Physics, Massachusetts Institute of Technology, Cambridge, Massachusetts 02139, USA}

\author{Benjamin E. Feldman}
\email{bef@stanford.edu}
\affiliation{Geballe Laboratory for Advanced Materials, Stanford, CA 94305, USA}
\affiliation{Department of Physics, Stanford University, Stanford, CA 94305, USA}
\affiliation{Stanford Institute for Materials and Energy Sciences, SLAC National Accelerator Laboratory, Menlo Park, CA 94025, USA}

\maketitle

\section{Materials and Methods}
\subsection{Sample fabrication}
The tWSe$_2$ device was fabricated using standard dry transfer techniques. Using a poly(bisphenol A carbonate) (PC)/polydimethylsiloxane (PDMS) stamp, we pick up a thin (15 nm thick) hexagonal boron nitride (hBN) flake, followed by the first half of the monolayer WSe$_2$ flake (exfoliated from HQ Graphene source), using the hBN to tear the flake in two, and then the second half rotated to a controlled angle of $1.5^\circ$. Separately, we prepare a stack with a bottom hBN (13 nm) and a graphite (5 nm) back gate, on which we deposit pre-patterned Cr/Pt contacts (2 nm / 8 nm). This is annealed at $\approx 300\ ^\circ$C for 8 hours to clean polymer and resist residues both before and after pre-patterning of contacts. The tWSe$_2$ stack is then dropped onto the pre-patterned contacts. We used standard e-beam lithography techniques to fabricate metallic contacts, as well as local ``contact-gates" over the Pt contacts, while leaving the rest of the sample ungated on top for access to the SET \cite{movva_high-mobility_2015,foutty_tunable_2023}. A second tWSe$_2$ device with lower twist angles ($\theta = 0.76^\circ$ to $\theta =1.11^\circ$) was fabricated with a similar process (Supplementary Sec. 7).

\subsection{SET Measurements}

The SET sensor was fabricated by evaporating aluminum onto a pulled quartz rod, with an estimated diameter at the apex of $ 50 - 100$ nm. The SET ``tip" is brought to about $50$ nm above the sample surface. Scanning SET measurements were performed in a  Unisoku USM 1300 scanning probe microscope with a customized microscope head. a.c. excitations (2-5 mV peak-to-peak amplitude) were applied to both sample and back gate at distinct frequencies between 200 and 400 Hz. We then measure inverse compressibility $\textrm{d}\mu/\textrm{d}n \propto I_{\textrm{BG}} / I_{\textrm{2D}}$ where $ I_{\textrm{BG}}$ and $I_{\textrm{2D}}$ are measurements of the SET current demodulated at respective frequencies of the back gate and sample excitations \cite{yu_correlated_2022}. Except where otherwise noted (Fig. 4 in the main text), a d.c. offset voltage $V_{\textrm{2D}}$ is applied to the sample to maintain the working point of the SET at its maximum sensitivity point within a Coulomb blockade oscillation fringe chosen to be near the ``flat-band'' condition where the tip does not gate the sample. This minimizes tip-induced doping and provides a direct measurement of $\mu(n)$ at d.c. timescales. Depending on measurement location, we measure a small ($<1\times 10^{-11}$ meV cm$^{2}$) difference in d$\mu$/d$n$ between a.c. and d.c. measurements, which we subtract from the a.c. in all data presented in the main text (see Supplementary Sec. 5 for further discussion). The contact gates are held at a large, negative voltage throughout the measurement to maintain good electrical contact across variable hole doping. All SET measurements are taken at $T = 330$ mK unless otherwise noted. Unless otherwise noted, all measurements in a magnetic field are taken with `positive' magnetic field pointing perpendicularly out of the plane of the device (towards the SET tip, as shown in Fig. 4 in the main text).

\subsection{Density and twist angle determination}
The approximate twist angle of the sample ($\sim 1.5^\circ$) is determined by the control over rotation in the stackng process. To ensure that the tWSe$_2$ is not overly strained or relaxed after picking it up, we perform piezoelectric force microscopy (PFM) using a Bruker Icon atomic force microscope during the fabrication process on the PC slide before setting the stack onto the prepatterned Pt contacts \cite{mcgilly_visualization_2020,bai_excitons_2020}. We present this in Supplementary Sec. 6. We find a relatively low-strain triangular moir\'e superlattice with lattice constant around $13$ nm, consistent with what we extract from SET measurements.

As described in the main text, there is some variation in local twist angle across the device. We use SET measurements to precisely determine local angle in a similar manner to standard techniques in magic-angle twisted bilayer graphene \cite{zondiner_cascade_2020,uri_mapping_2020}. From the slopes of the Hofstadter states, we can accurately measure the sample capacitance and convert between back gate voltage $\rm{V}_{\rm{BG}}$ and carrier density $n$. We then use the integer gaps we measure (e.g. at $\nu = -1,-2$, and $-3$) and/or the corresponding Hofstadter intercepts (in cases where there is no zero-field gap) to determine the density corresponding to filling one hole per moir\'e unit cell. From this density $n_{\rm{muc}}$, we convert to twist angle $\theta$ via $1/n_{\rm{muc}} = \frac{\sqrt{3}a_{\rm{WSe}_2}^2}{4-4\cos\theta}$ and subsequently moir\'e wavelength via $\lambda_m = \frac{a_{\rm{WSe}_2}}{2\sin(\theta/2)}$, where $a_{\rm{WSe}_2}=0.328$ nm.

\subsection{Extraction of gap sizes}
The thermodynamic gap sizes shown in the main text are given by the size of the step in the chemical potential $\mu(n)$. Practically, this is extracted by numerically integrating $\textrm{d}\mu/\textrm{d}n$ across the gap. To accurately measure the gap on top of the widespread negative compressibility coming from long-range interactions at low density, we subtract a small background before integrating, analogous to Refs. \cite{eisenstein_compressibility_1994,kometter_hofstadter_2022}. Generally, this background is taken from averaging the value of $\textrm{d}\mu/\textrm{d}n$ on either side of the gap. This background is measured sufficiently far (in density) from the gap so that chemical potential behavior immediately adjacent to the gap, for example enhanced negative compressibility, is not included in the background. In a few cases in which gaps are particularly close together (for example, $C=+1$ gaps and $C=0$ gaps with $s = -1$ between $B = 0.5$ and $B = 1.5$ T), the background is just taken from a single side of the gap, so that the behavior of one gap does not affect the extracted value of the other. In general, disorder can reduce the size of measured gaps. In the main text (Figs. 2-3), we present gaps from data measured at optimized locations that have low twist angle variability in order to minimize the effects of disorder in comparisons.

\subsection{Electric field tuning with SET}
As mentioned in the main text, if the d.c. sample voltage $V_{\textrm{2D}}$ is tuned away from the ``flat-band" condition compensating for the work function difference between sample and tip, the tip will locally gate the sample. We can model this doping by treating the tip as one side of a parallel plate capacitor: $D_{\textrm{eff}}= \frac{1}{2\epsilon_0}(C_t(0-(V_{2D}-V_{fb})) - C_b (V_{BG}-V_{2D}-V_0))$, where $C_{b(t)}$ is the back (top) gate capacitance, $V_{2D}$ and $V_{BG}$ are the d.c. voltages applied to sample and back gate, $V_{fb}$ is the ``flat-band" voltage at which the tip and sample are work-function-compensated, and $V_0$ is the voltage at which back gate and sample are work-function-compensated (equivalent to the voltage of the WSe$_{2}$ band edge). While $C_t$ will depend on the height of the tip, we experimentally extract $C_t \approx 0.045C_b$ for the measurements shown in the main text based on the shifts of constant-density features in the $V_{BG}-V_{2D}$ plane. Data shown in Fig. 4c-d in the main text is taken by initially fixing $V_{2D}$ and then sweeping $V_{BG}$, feeding back on the value of $V_{2D}$ while data is taken to maintain the same position along the SET Coulomb blockade oscillation. From this data, we apply the transformation to convert to $n$ and $D_{\textrm{eff}}$ from $V_{2D}$ and $V_{BG}$. 

\subsection{Hofstadter spectrum calculation}
We compute the Hofstadter spectrum for the continuum model for TMD homobilayers~\cite{wu_topological_2019},
with continuum model parameters extracted from ab initio calculations of Ref~\cite{devakul_magic_2021}: $(V,w,\psi)=(9\mathrm{\ meV},18\mathrm{\ meV},128^\circ)$.
We use an effective mass $m=0.3m_e$, where $m_e$ is the electron mass, and a WSe$_2$ lattice constant of $a_0 = 3.317$\AA.
The calculation is performed at a twist angle $\theta=1.2^\circ$ and displacement field is modeled as an interlayer potential $\mathrm{diag}(\frac{\Delta}{2},-\frac{\Delta}{2})$ with $\Delta=4$ meV.  
The continuum model is described in more detail in Supplementary Sec. 2-3.
The spectrum for the $K$ valley bands features a topological first band with $C=1$ that is close to a displacement field tuned band inversion with the second band.

The finite magnetic field is incorporated by minimal substitution $\vec{p}\rightarrow\vec{\pi}=\vec{p}-\vec{A}$ with the symmetric gauge vector potential $A=\frac{B}{2}(x\hat{y}-y\hat{x})$.
At flux per unit cell $B=\frac{p}{2q A_{uc}}$, where $A_{uc}$ is the moir\'e unit cell area, the Hamiltonian is diagonalized in the Landau level basis with a large cutoff $N_{\mathrm{max}}=\lfloor100q/p\rfloor$, using the method described in detail in Ref~\cite{kometter_hofstadter_2022}.
The resulting Hofstadter spectrum shown in Fig.~2d in the main text is computed for the $K$ valley bands.

To compute the Wannier plot in Fig.~2e in the main text, we first apply a Gaussian broadening of width $\sigma=0.5$ meV to the calculated density of states.  
The filling factor $\nu$ is obtained by integrating the broadened density of states from the charge neutrality point, assuming full valley polarization.

\section{Supplementary Text}

\subsection{1. Topological gaps around $B = 0$}

In Supplementary Fig. \ref{fig:Hysteresis}, we present high-resolution measurements of the inverse electronic compressibility d$\mu$/d$n$ near moir\'e filling factors $\nu = -1$ and $\nu = -3$ at low magnetic field $B$ in the same location as in Fig. 1-2 of the main text (with twist angle $\theta = 1.23^\circ$). These states continue to be sloped all the way down to $B = 0$. Upon switching the field polarity (to $B<0$), the slope reverses direction, as a $C = -1$ state is favored by an opposite magnetic field. For these measurements, the density axis is the ``fast" axis, and is swept back and forth in a small range around each gap before changing the magnetic field. Under these conditions, we do not resolve any hysteresis; because our measurement is only sensitive to the sign of magnetization via the slope (i.e. we require multiple points to take a numerical derivative), this yields an upper bound for the coercive field of $<10$ mT. Very small coercive fields of this order of magnitude have been reported in similar systems \cite{li_quantum_2021,tao_valley-coherent_2022}. The doubling in the state at $\nu = -3$ likely reflects a small amount of potential and/or twist angle disorder in this location. The persistence of sloped gaps all the way to zero magnetic field with a resolution of 5 mT unambiguously shows that these are true quantum anomalous Hall states and are distinct from Landau levels.

\subsection{2. Continuum model Hartree-Fock analysis}
In this section, we theoretically study the continuum model for twisted WSe$_2$ via self-consistent Hartree-Fock (HF) analysis. Our starting point is the single-particle continuum model Hamiltonian, written here as a matrix in layer pseudospin \cite{wu_topological_2019, devakul_magic_2021}
\begin{align}
    \mathcal{H}_{\uparrow} = \begin{pmatrix}
    \frac{\hbar^2(\bm k - \bm \kappa_+)^2}{2m} + V_1(\bm{r}) & t(\bm{r}) \\
    t^{\dag}(\bm{r}) & \frac{\hbar^2(\bm k - \bm \kappa_-)^2}{2m} + V_2(\bm{r})
    \end{pmatrix}.
    \label{eq:Hupcont}
\end{align}
Due to strong spin-orbit coupling, spin and valley are ``locked" at low energy into a single $SU(2)$ degree of freedom that we refer to simply as spin. $V_l(\bm{r})$ is an intra-layer moiré potential and $t(\bm{r})$ is an inter-layer tunneling. We choose our moiré reciprocal lattice vectors to be $\bm{g}_i=\frac{4\pi}{\sqrt{3}a_M}(\cos\frac{\pi(i-1)}{3},\sin\frac{\pi(i-1)}{3})$. The vectors $\bm \kappa_+ = \frac{\bm g_1 +\bm g_6}{3}$, $\bm \kappa_- = \frac{\bm g_1 +\bm g_2}{3}$ account for the momentum space displacement of the two layers' Brillouin zone corners upon rotation. Their lowest-harmonic Fourier expansions are constrained by symmetries to the form
\begin{align}
    V_l(\bm{r})=-2V\sum_{i=1,3,5}\cos(\bm{g}_i\cdot\bm{r}+\phi_l) + (-1)^l \frac{\Delta}{2}
\end{align}
with $\phi_2=-\phi_1$ and
\begin{align}
    t(\bm{r}) = w(1+e^{i\bm{g}_5\cdot\bm{r}}+e^{i\bm{g}_6\cdot{\bm{r}}}).
\end{align}
Here $\Delta$ is an inter-layer voltage difference corresponding to a displacement field.
We remark that Eq~\ref{eq:Hupcont} describes the physics of holes (i.e. the sign of the kinetic term is positive), although we will typically plot the electron band structures, which are related by a sign flip.
We use the continuum model parameters extracted from density functional theory calculations reported in Ref. \cite{devakul_magic_2021}: $V=9.0\text{ meV}$, $w=-18\text{ meV}$, and $\phi=128^{\circ}$, along with effective mass $m=0.3m_e$ ($m_e$ is the electron mass) and WSe$_2$ lattice constant $a_0=3.317$ \AA. The moir\'e period is $a_M\equiv \frac{a_0}{2\sin(\theta/2)} \approx \frac{a_0}{\theta}$. Time reversal symmetry ensures that $\mathcal{H}_{\downarrow}$ is the time reversal conjugate of $\mathcal{H}_{\uparrow}$. Explicitly,
\begin{align}
    \mathcal{H}_{\downarrow} = \begin{pmatrix}
    \frac{\hbar^2(\bm k + \bm \kappa_+)^2}{2m} + V_1(\bm{r}) & t^\dag (\bm{r}) \\
    t(\bm{r}) & \frac{\hbar^2(\bm k + \bm \kappa_-)^2}{2m} + V_2(\bm{r})
    \end{pmatrix}.
\end{align}

We perform the HF calculations in a plane wave basis following the methodology described in the supplements of Refs. \cite{reddy_moire_2023, kometter_hofstadter_2022}, with the straightforward addition of layer pseudospin. We use an interaction that accounts for the presence of a single metallic gate and finite interlayer separation:
\begin{align}
    U_{ll'}(\bm q) = \frac{e^2}{2\epsilon\epsilon_0}  \frac{1-e^{-2d_Gq}}{q} \times \begin{cases}
    1 &\text{$l=l'$}  \\
    e^{-d_{I}q} &\text{$l\neq l'$}. 
    \end{cases}
\end{align}
Here $q=|\bm{q}|$, $d_I=7$ \AA\ is the out-of-plane separation between the two TMD monolayers, and $d_G=130$ \AA\  is the out-of-plane separation between the center of the TMD bilayer and the nearby surface of the metallic gate. This expression assumes $d_I\ll d_G$.

We perform two types of HF calculations: ``unrestricted" (UHF), in which the single-particle density matrix is forced to be diagonal in spin, and ``generalized" (GHF) in which the single-particle density matrix is allowed to have non-vanishing off-diagonal-in spin components. As we will see, the latter is essential to capture states with in-plane magnetic order.
We use a momentum space cutoff of $4.5|\mathbf{g}_1|$, which corresponds to keeping 73 reciprocal lattice vectors, and a momentum space mesh of $15\times15$ points.
We start the self-consistent calculation with non-interacting density matrices and small random noise.
For the GHF calculations, we apply a strong in-plane valley Zeeman field $-\frac{h}{2}\sum \sigma_x$ with $h=10$ meV, where $\sigma_x$ is the spin(+valley) Pauli matrix, for the first 5 cycles in order to bias towards an intervalley coherent state.

Figure \ref{fig:HFFig}a-c shows the UHF phase diagram obtained at integer filling factors $n_{tot}=n_{\uparrow}+n_{\downarrow}=|\nu|$ for various spin(+valley) polarizations $(n_{\uparrow}, n_{\downarrow})$ as a function of the interaction strength $\epsilon^{-1}$  and twist angle $\theta$. 
At $n_{tot}=1$ we find regions of non-trivial and trivial topology (cyan and yellow dots, respectively, see legend for Chern numbers to the right).
At $n_{tot}=2$, we see regions of ferromagnetism $(n_{\uparrow},n_{\downarrow})=(2,0)$ with Chern numbers $0$ and $2$, as well as regions with zero total magnetization $(n_{\uparrow},n_{\downarrow})=(1,1)$.
We have explicitly checked that the latter region corresponds to a quantum spin Hall (QSH) insulator in which the filled bands carry opposite spin-contrasting Chern numbers.
At $n_{tot}=3$, the state is partially polarized with $(n_{\uparrow},n_{\downarrow})=(2,1)$ and we find regions of parameter space with a range of Chern numbers $C=-1,0,1,2$.
In a large portion of parameter space, we find that the $n_{tot}=1$ and the $n_{tot}=3$ state have the same Chern number sign $C=1$, while the $n_{tot}=2$ state is a QSH insulator.
Assuming that these are the spin configurations present in the experiment at integer $\nu$ (see below), this is consistent with the experimental observations of the same sign of the Chern number at $\nu=-1$ and $\nu=-3$, and a $C=0$ $\nu=-2$ state.

Here, we discuss in more detail the evidence that the Chern numbers of the first two moir\'e bands have the same sign for a given valley. First, the spin $g$-factor in TMD moir\'e valence bands is significantly larger than the calculated orbital effects of the moir\'e bands (Supplementary Sec. 3). Thus, for a given direction of applied magnetic field, the preferred spin/valley polarization should be the same at $\nu = -1$ and $\nu = -3$. Additionally, measurements of the inverse electronic compressibility are highly sensitive to (first-order) spin transitions \cite{yu_correlated_2022}, and we indeed observe sharp negative compressibility in an applied magnetic field which indicates such phase transitions (Fig. 2 in the main text). At $\nu = -1$, we observe no spin/valley transitions along the $C = +1$ gapped state, suggesting that the insulator at $B = 0$ has the same spin configuration as that favored by the Zeeman effect at high fields. At $\nu = -2$, we observe a single sharp transition around $B = 2$ T, whose most natural interpretation is a transition between a spin configuration of $(n_{\uparrow}, n_{\downarrow}) = (1,1)$ to $(n_{\uparrow}, n_{\downarrow}) = (2,0)$, where $n_{\uparrow(\downarrow)}$  is the number of particles per moir\'e unit cell in the the spin/valley direction favored (disfavored) by the magnetic field (similar to the moir\'e band crossing discussed in \cite{kometter_hofstadter_2022}). At $\nu = -3$, we also observe a single transition, at a slightly higher field. Because we only observe one (rather than two) transitions, this is consistent with a single spin flip from $(n_{\uparrow}, n_{\downarrow}) = (2,1)$ to $(n_{\uparrow}, n_{\downarrow}) = (3,0)$. While our data do not explicitly rule out intermediate spin phases and additional phase transitions beyond the field range available in experiment, this is unlikely due to the large spin $g$-factors discussed above.

Next, we examine the states with intervalley cohererence (IVC) at $n_{tot}=1$, using GHF.  
As our calculation assumes translation invariance, we can only capture IVC states that are periodic on the moir\'e scale (i.e. XY ferromagnetic (FM) states).
To describe XY antiferromagnetic (AFM) states with $120^\circ$ order and an enlarged $\sqrt{3}\times\sqrt{3}$ unit cell, we repeat the GHF calculation for various choices of spin-contrasting shifts $\zeta=1,2$.  
We refer to the original (unshifted) configuration as $\zeta=0$.
This simply involves a uniform shift of the $\kappa_{\pm}$ points in the moir\'e Brillouin zone, as illustrated in Fig.~\ref{fig:HFFig}d for the two options $\zeta=1,2$.
This transformation maps the 120$^\circ$ XY AFM state with the appropriate chirality into an XY FM state that can be described with our translation invariant GHF ansatz~\cite{devakul_quantum_2022}.

Figure~\ref{fig:HFFig}e shows the HF energy as a function of $\theta$ for a fixed $\Delta$ and $\epsilon$.  
A (topologically trivial) IVC state with $\zeta=1$ becomes energetically favorable at high twist angles.  
We find that GHF fails to converge in some regions of parameter space (e.g. GHF $\zeta=0$ for $\theta>2^\circ$ in Fig.~\ref{fig:HFFig}e).
Finally, Fig.~\ref{fig:HFFig}f shows the $n_{tot}=1$ phase diagram including the IVC state as a function of $\theta$ and layer potential $\Delta$.  
While qualitatively similar to the experimental $\nu=-1$ phase diagram in the main text (a valley polarized (VP) quantum anomalous Hall (QAH) state appears flanked by topologically trivial phases), the theoretical QAH phase appears in a much wider range of angles than found experimentally.
This is likely due to uncertainty in parameter choice and inadequacy of the simple continuum model (which only includes first harmonic potential and tunneling terms) in describing low twist angle WSe$_2$ where lattice relaxation effects are strong \cite{shabani_deep_2021,van_winkle_rotational_2023} (see further discussion below). Another contributor is that HF will tend to better describe the fully valley polarized states in comparison to the IVC state which, by analogy with XY AFM state of the triangular lattice Hubbard model, has stronger quantum fluctuations.  
Thus, HF will tend to overestimate the extent of the VP phase (which only has a small energy difference $\sim 0.5$ meV, Fig.~\ref{fig:HFFig}e, compared to the IVC state).

\subsubsection{Relaxation effects on electronic structure}

In this section, we study the qualitative effect of lattice relaxation, which becomes significant at small twist angles, on the theoretical band structure.
The out-of-plane lattice relaxation is incorporated in the interlayer tunneling and potential terms obtained from the shift method~\cite{wu_topological_2019,devakul_magic_2021}, hence we focus on the in-plane lattice relaxation.
At small non-zero twist angles, $\theta\approx 0$, in-plane lattice relaxation leads to the formation of triangular domains of energetically favorable MX/XM regions, separated by a network of domain walls~\cite{shabani_deep_2021,van_winkle_rotational_2023,arnold_relaxation_2023,vitale_flat_2021}.
At small angles, these domain walls have a characteristic width $\xi$.  
We remark that lattice relaxation is very different in parallel (our case, $\theta\approx 0$) versus anti-parallel ($\theta\approx 60^\circ$) twisted WSe$_2$ homobilayers.  
While the former relaxes into large triangular domains, the latter relaxes into hexagonal domains \cite{li_lattice_2021,enaldiev_stacking_2020}.

We introduce an extension of the continuum model for TMD homobilayers that incorporates the main effect of in-plane lattice relaxation, with single tuning parameter $\xi$, the MX/XM domain wall width.  
This model allows us to smoothly tune between a heavily reconstructed ($\xi\rightarrow 0$) limit, and the rigid ($\xi\rightarrow\infty$) limit which
 reduces to Eq.~\ref{eq:Hupcont}.

 We treat the relaxed structure using the continuum approach.  
 We consider two layers labeled by $l=1,2$, twisted at angles $\theta_l = (\theta/2,-\theta/2)$, where $\theta$ is a small angle close to zero.
 At a position $\vec{r}$, the lattice in the relaxed structure in layer $l$ appears shifted by a vector $\vec{U}_l(\vec{r})=\theta_l \hat{z}\times\vec{r}+\vec{u}_l(\vec{r})$, where the first term is due to the rigid twist, and $\vec{u}_l(\vec{r})$ is a moir\'e periodic contribution due to lattice relaxation.
 
 We consider a simple analytic form for the relaxed shifts $\vec{u}_l(\vec{r})$ which captures the main features found by direct minimization of elastic potential energy.
Assuming that the relaxed structure maintains the same symmetry as the unrelaxed structure, that there is no biaxial compression $\nabla\cdot \vec{u}_l(\vec{r})=0$, and that the domain walls are $\tanh$-shaped in the small $\theta$ limit, we are led to the following expression
\begin{equation}
\vec{u}_l(\vec{r}) = \frac{-\theta_l\pi\xi}{\sqrt{3}}\sum_{j=1,3,5}\frac{\vec{g}_j\times\hat{z}}{|\vec{g}_j|} \sum_{n=1}^{\infty}\mathrm{csch}\left(\frac{\xi\pi^2 n}{a_M}\right)\mathrm{sin}(n \vec{g}_j\cdot\vec{r})
\end{equation}
where $\vec{g}_j=\frac{4\pi}{\sqrt{3}a_M}(\cos(\frac{j-1}{3}\pi),\sin(\frac{j-1}{3}\pi))$ are the moir\'e reciprocal lattice vectors, and $a_M = a_0/\theta$, where $a_0=0.3317$ nm is the atomic lattice constant.
This relaxed structure is moir\'e periodic and maintains the $C_{3z}$ and $C_{2y}$ symmetry.  
The local twist angle is given by $\theta_l(\vec{r})=\frac{1}{2}(\nabla\times\vec{U}_l(\vec{r}))_z$,
\begin{equation}
\theta_l(\vec{r}) = \frac{\theta_l}{3}\sum_{j=1,3,5} \sum_{n=0}^{\infty}\frac{\pi^2\xi n}{a_M} \mathrm{csch}\left(\frac{\xi\pi^2 n}{a_M}\right)\mathrm{cos}(n\vec{g}_j\cdot\vec{r})
\end{equation}
which is illustrated in Fig.~\ref{fig:RelaxationFig}a for $\theta=1.2^\circ$ and $\xi=3$nm.
As can be seen, for small $\xi$, the resulting structure relaxes to large regions that are locally untwisted, $\theta_l(\vec{r})\approx 0$, separated by domain walls where the local twist angle is non-zero.
By construction, across the domain walls, the local atomic alignment $\vec{U}_1(\vec{r})-\vec{U}_2(\vec{r})$ jumps from MX to XM stacking as $\sim \tanh(x/\xi)$.
We remark that while the exact analytic form for the domain walls in the small angle limit is not precisely of this form \cite{nam_lattice_2017,gao_symmetry_2022}, we expect the $\tanh$ form used here to capture the essential physics.

To determine the effect on electronic structure, we derive the coupling of the shift field to the continuum model of Eq.~\ref{eq:Hupcont}.
We work in the gauge where the kinetic terms are centered about $\frac{|\vec{k}|^2}{2m}$ for both layers, such that the Bloch wavefunction satisfies $\psi_{k,l}(\vec{r}+\vec{a}_M)=e^{i (\vec{k}-\vec{K}_l)\cdot\vec{a}_M}\psi_{k,l}(\vec{r})$.  
The potential and tunneling terms are derived from the local-shift-dependent terms~\cite{wu_topological_2019}, given by 
\begin{eqnarray}
\mathrm{V}_1(\vec{d}) = 2V\sum_{i=1,3,5}\cos (\vec{G}_i\cdot\vec{d}+\phi)+\frac{\Delta}{2}\\
\mathrm{V}_2(\vec{d}) = 2V\sum_{i=1,3,5}\cos (\vec{G}_i\cdot\vec{d}-\phi)-\frac{\Delta}{2}\\
\mathrm{T}(\vec{d}) = w e^{i\frac{\vec{G}_2+\vec{G}_3}{3}\cdot\vec{d}}(1+e^{-i\vec{G}_2\cdot\vec{d}}+e^{-i\vec{G}_3\cdot\vec{d}})
\end{eqnarray}
where $\vec{G}_i=\frac{4\pi}{\sqrt{3} a_0}(-\sin(\pi\frac{j-1}{3}),\cos(\pi\frac{j-1}{3}))$ are the atomic reciprocal lattice vectors.
The corresponding moir\'e potential and tunneling terms are obtained as 
\begin{eqnarray}
V_l(\vec{r}) = \mathrm{V}_l(\vec{U}_1(\vec{r})-\vec{U}_2(\vec{r}))\\
T(\vec{r}) = \mathrm{T}(\vec{U}_1(\vec{r})-\vec{U}_2(\vec{r}))
\end{eqnarray}
which reduce to the usual continuum model, Eq.~\ref{eq:Hupcont}, in the absence of relaxation, $\vec{u}_l(\vec{r})=0$.
We plot the real space potential and tunneling terms in Fig.~\ref{fig:RelaxationFig}b-d for $\theta=1.2^\circ$ and $\xi=3$ nm, which shows the expected result that the potentials are constant in the MX and XM domains, and reproduce their extracted value within the domains as obtained by shift-method DFT~\cite{devakul_magic_2021}.
To incorporate these into the continuum model, we take the numerical Fourier transform of the real-space functions.

Next, we consider the coupling of the shift field to the kinetic energy.
It is known that for $K$ valley materials, including graphene, the shift field couples to lowest order as a spatially-dependent valley-contrasting gauge field (called a pseudomagnetic gauge field) \cite{suzuura_phonons_2002,pereira_strain_2009,nam_lattice_2017,xie_valley-polarized_2022}.  
To derive this, we model the band edge of monolayer WSe$_2$ as arising from a honeycomb lattice of M and X sublattices, with a sublattice potential mass term.
The hopping integrals $t(\delta)$ along a bond in the $i$ direction is taken to depend only on the interatomic spacing $\delta$, which is $\delta=b\equiv a_0/\sqrt{3}$ in the unrelaxed case.  
In the presence of uniform strain, i.e. a uniform $u_{ij}=\partial_i u_j(\vec{r})$, the bond lengths are modified and the hopping integral becomes 
\begin{equation}
t( b - b u_ii) \approx t_0 - t_0 \beta u_{ii}
\end{equation}
where $t_0\equiv t(b)$ and $\beta\equiv \left.-\frac{d\log t(\delta)}{d\log \delta}\right|_{\delta=b}$ is a value of $O(1)$.
\begin{equation}
H_{tb}(\vec{k}) = s\sigma^z + \left(\sigma^+ t_0\sum_{i=1,2,3}(1-\beta u_{ii})e^{-i \vec{d}_i \cdot\vec{k}} + h.c.\right)
\end{equation}
where $\sigma$ are Pauli matrices,
$s$ is the sublattice mass term, 
$\vec{d}_1=-\vec{b}_1+2\vec{b}_2,
\vec{d}_2=-\vec{b}_1-\vec{b}_2,
\vec{d}_3=2\vec{b}_1-\vec{b}_2$ are the three hopping vectors where $\vec{b}_1=b(1,0)$ and $\vec{b}_2=b(1/2,\sqrt{3}/2)$.
In terms of the Cartesian components, $u_{11}=u_{yy}$, $u_{22}=\frac{3u_{xx}+u_{yy}+\sqrt{3}(u_{xy}+u_{yx})}{4}$, and $u_{22}=\frac{3u_{xx}+u_{yy}-\sqrt{3}(u_{xy}+u_{yx})}{4}$.
Expanding the kinetic energy about $K$, $k=K+p$, we find the effective dispersion, to first order in $\beta$ and second order in $p$,
\begin{equation}
E(\vec{p})\approx \frac{1}{2m}\left(|\vec{p}|^2 + \beta\left[
\frac{\sqrt{3}}{b}\vec{p}\cdot\vec{w} +|\vec{p}|^2 w_0 +\frac{1}{2}(|\vec{p}|^2 w_x - 2 p_x p_y w_y)
\right]\right)
\end{equation}
where $\vec{w}=(w_x,w_y), w_x=u_{xx}-u_{yy}, w_y=-u_{xy}-u_{yx}$, and $w_0=u_{xx}+u_{yy}$, and we have identified the mass $m=4s/(3t^2a^2)$.  
The first term is the usual parabolic dispersion.
In the square brackets, the corrections can be interpreted as
the pseudomagnetic gauge field, an effective mass correction, and a mass anisotropy term.
For our particular choice of $u$, $w_0=\nabla\cdot\vec{u}_l(r)=0$.
To incorporate these terms into the continuum model, we incorporate the spatial dependence of $\vec{w}(\vec{r})$ and $w_0(\vec{r})$, and take the symmetric ordering, i.e. replacing $p_x w_x(\vec{r})\rightarrow \frac{1}{2}\left\{p_x,w_x(\vec{r})\right\}$.
We show $\vec{w}(\vec{r})$ in Figure~\ref{fig:RelaxationFig}e,f.

Figure~\ref{fig:RelaxationBandsFig}a shows the moir\'e band structure computed for the continuum model at the same parameters used for the calculation of the Hofstadter states in the main text, for $\xi=2$ nm, and $\beta=2$.  
We show the bands with only the relaxed potential and tunneling terms, only the relaxed kinetic terms, and both.  
As can be observed, the relaxed potential and tunneling terms have the largest effect, while the kinetic term results in fairly small corrections.
Figure~\ref{fig:RelaxationBandsFig}b,c show the band structure at $\theta=1.2^\circ$ and $\theta=1.4^\circ$, for a few choices of $\xi$.
While relaxation quantitatively affects the dispersions and band gaps, there do not appear to be any major changes, such as gap closings,  for reasonable domain wall widths ($\xi\sim 3$ nm) \cite{enaldiev_stacking_2020}.

\subsection{3. Band structure and orbital magnetization}
In this section, we show the band structure for the model used to compute the Hofstadter spectrum in the main text, and demonstrate the topological phase transition tuned by displacement field.  
In Fig.~\ref{fig:MagnetizationBandsFig}a, we show the continuum model  (Eq.~\ref{eq:Hupcont}) band structure at $\Delta=4,6,8$ meV, through which a topological band inversion occurs.  
The Hofstadter spectrum calculated in the main text correspond to $\Delta=4$ meV, where the first band is topological.  

Next, we calculate the orbital magnetization assuming full valley polarization as a function of $\mu$.  
This is given by the formula \cite{xiao_berry_2010}
\begin{equation}
M_{\mathrm{orb}}=-\frac{e}{\hbar}\sum_n\int \frac{d^2k}{4\pi^2} f(E_n(k)-\mu) \sum_{m\neq n}\mathrm{Im}\frac{\bra{n(k)}\partial_{k_x}H\ket{m(k)}\bra{m(k)}\partial_{k_y}H\ket{n(k)}}{(E_n(k)-E_m(k))^2}\left[(E_n(k)-E_m(k))+2(\mu-E_n(k))\right]
\end{equation}
which is plotted in Fig.~\ref{fig:MagnetizationBandsFig}b.  
We find that the orbital magnetization is of order $\sim 1-2\mu_B$ per moir\'e unit cell.  
The orbital Zeeman energy is therefore much smaller than the expected spin-Zeeman energy (with spin $g$-factors $g\sim 10$ \cite{wang_valley-_2017,zhao_realization_2022}).

While the single-particle Hofstadter spectrum shown in Fig. 2 in the main text qualitatively matches with experiments near $\nu = -1$, the experimental behavior at higher doping is more complicated. At high fields, we expect the gap at $\nu = -3$ to be a fully spin/valley polarized gap driven by interactions. Because of that, one might expect the Landau levels upon doping away from this gap to be singly degenerate, so that the Landau fan would include states with every integer Chern number \cite{wang_correlated_2020,kometter_hofstadter_2022}, rather than having any underlying degeneracy that would change the sequence of Landau levels. However, the pattern that we observe is more complicated, as some Hofstadter states are much weaker than others. For example, $(C,s) = (+2,-3)$ is much weaker than $(+1,-3)$ or $(+3,-3)$. In flat electronic bands, the exact details of which Hofstadter states are observed can be quite complicated due to the small energy scales of single-particle Hofstadter gaps relative to interactions \cite{kometter_hofstadter_2022}.



\subsection{4. $D$-field tuning at other angles} 
In this section, we provide supporting data used to determine the phase diagram in Fig. 4g. We present measurements in Fig. \ref{fig:1.23_1.25deg} that are analogous to those in Fig. 4c-f at two independent locations with different twist angles. Data from the same location as Figs. 1-2, with a twist angle of $\theta=1.23^\circ$, is plotted in Fig. \ref{fig:1.23_1.25deg}a-d, and Fig. \ref{fig:1.23_1.25deg}e-f shows data measured at a location with $\theta = 1.25^\circ$. The critical displacement fields $D_c$ are taken to coincide with the minima of the $\nu = -1$ gap vs. $D_{\rm{eff}}$ for $B=0$. Different twist angles yield different values for $D_c$, around $68$ mV/nm at $\theta = 1.23^\circ$ (Fig. \ref{fig:1.23_1.25deg}a) and $48$ mV/nm at $\theta = 1.25^\circ$ (Fig. \ref{fig:1.23_1.25deg}e). While the extracted values for $D_c$ are approximately the same if we consider the change at filling $\nu = -1$ in a finite magnetic field between states centered at $C = +1$ and $C = 0$ (Fig. \ref{fig:1.23_1.25deg}b,f), there is occasionally a small discrepancy in the displacement field of $\sim 5$ mV/nm. (This is most apparent in Fig. \ref{fig:1.23_1.25deg}a-b). We believe that this discrepancy likely stems from the overall smearing of the transition due to the inhomogeneous displacement field applied by the SET tip, and is likely not indicative of a shift of the critical displacement field in an applied magnetic field (our measurement provides an upper bound to such a shift). If the spin-valley character of the two states above and below $D_c$ are different, one might expect that $D_c$ should have some measurable dependence in an applied magnetic field, while an identical $D_c$ would suggest that the states share the same spin-valley ordering. Our experiments provide a constraint that suggests the latter case and motivate future experiments to more precisely measure $D_c$ and its magnetic field dependence to determine the spin ordering across the phase diagram.

In Fig. \ref{fig:1.185deg}, we present data from the same $\theta = 1.19^\circ$ location as shown in the second from the top panel of Fig. 3a, which exhibited trivial behavior at $\nu = -1$ when the tip did not provide any additional doping. By lowering the displacement field with the SET tip, we locally tune the $\nu = -1$ state into a topological $C = +1$ phase, as identified both by the shift in density of the state at finite field (Fig. \ref{fig:1.185deg}a), as well as the slope as a function magnetic field (Fig. \ref{fig:1.185deg}b). Using the tip to increase the applied displacement field maintains trivial behavior (Fig.~\ref{fig:1.185deg}c).

\subsection{5. Background subtraction from a.c. charging}
As described in the Methods section, the single-electron transistor (SET) simultaneously and independently measures the chemical potential $\mu(n)$ and the inverse electronic compressibility d$\mu$/d$n$ on d.c. and a.c. timescales, respectively. In the case where the RC time constant to charging the sample is comparable to the measurement frequency, the a.c. measurement of d$\mu$/d$n$ can become artificially enhanced. This is not a major problem in our experiment, as the contact gates are able to lower the Schottky barriers sufficiently to charge the sample. However, we consistently measure a small density-independent enhancement $<1\times 10^{-11}$ meV cm$^{2}$ between the integral of the a.c. measured d$\mu$/d$n$ and the derivative of the d.c. measured $\mu(n)$. This background can vary as a function of spatial position, and it generally gets worse the further that we get from the contacts. We account for this small enhancement by subtracting a constant from the measured d$\mu$/d$n$ to minimize the least squared difference between the a.c. and d.c. measurements. We present an example of this in Fig. \ref{fig:Background}. All data in the main text have had this spurious ``background" subtracted. This helps with side-by-side comparisons of d$\mu$/d$n$ measurements, such as in Fig. 3a, even when they are taken in far-separated locations. However, it does not affect any of the qualitative findings discussed in the manuscript. Additionally,  because a local (in density) background d$\mu$/d$n$ is subtracted in order to integrate the measured gap sizes (Methods), the subtraction described here does not quantitatively affect any measured gap sizes.

\subsection{6. PFM Characterization}
In this section, we provide further details of the piezoelectric force microscopy (PFM) characterization that we performed during the stacking process. In Fig. \ref{fig:PFM}, we present PFM data taken at a central location in the sample before encapsulation, when the twisted WSe$_2$ is supported by hBN and the PC stamp (Methods). As has been previously shown, the sample can relax after setting it down, so this may not correspond to the exact configuration in the final device \cite{bai_excitons_2020}. However, it still provides a rough estimate of the twist angle that is more accurate than that based on the targeted rotation during stacking. We measure the moir\'e wavelength $\lambda_m$ along the three principle directions via fast-Fourier-transforming as $13.5$, $13.2$, and $12.5$ nm. The small discrepancy from a perfectly triangular lattice can be described using previously developed models for uniaxial strain \cite{kerelsky_maximized_2019,zhang_flat_2020}. From these numerical models, we estimate a twist angle of $\theta = 1.44^\circ$, and a uniaxial strain of $\varepsilon=0.20 \%$ in a direction $\theta_S = 23.4^\circ$ away from the axis with $\lambda_m = 13.5$ nm. We note that this twist angle agrees quite closely to that which we estimate from the moir\'e unit cell area measured via SET measurements in a similar region of the sample (near location 5 in Supplementary Fig. \ref{fig:DeviceCharacterization}).

\subsection{7. Measurements of a second device}
All data presented in the main text and prior sections are from a single device. Within a single device, each measurement location would be expected to have different underlying disorder and strain configurations, as well as (often) distinct twist angles. Therefore, spatially resolved measurements should be interpreted as data from a large number of independent `samples', without requiring fabrication of multiple physical devices. Additionally, we have measured a second device (detailed below) with local twist angles that vary from $\theta = 0.76^\circ$ to $\theta = 1.11^\circ$. The corresponding data extends the overall twist angle range that we characterize and continues the trends we observed in the first device. We observe no indications of any nontrivial topological gaps in the angle range studied in the second sample, consistent with the trivial gaps observed below $\theta = 1.2^\circ$ in the first sample. 

In Fig. \ref{fig:SampleB}, we present data from the second device, which we refer to as Sample B. Due to worse electrical contacts, AC measurements of d$\mu$/d$n$ are only possible for hole densities above $0.9\times 10^{12}$ cm$^{-2}$ in this device. In Fig. \ref{fig:SampleB}a, we show inverse electronic compressibility at the highest twist angle measured, $\theta = 1.11^\circ$, where we observe a thermodynamic gap at $\nu = -3$ as well as a number of Hofstadter states emanating from $s=-3$ and $s=-4$. In Fig. \ref{fig:SampleB}b, we show data in a comparable angle from `Sample A’ studied in the main text (data from Fig. 3a in the main text), highlighting the similar pattern of incompressible features with red arrows. At this twist angle (the only one where we can make a direct comparison), measurements from the distinct samples qualitatively agree with one another. In Fig. \ref{fig:SampleB}c-d, we show data from lower twist angles in Sample B. As the twist angle decreases, gaps at higher integers become much stronger, appearing at $\nu = -4$ around $\theta = 1^\circ$ and $\nu = -5,-6$ at $\theta = 0.76^\circ$. This continues the trend observed in Sample A that more integer gaps are observed at lower twist angles, and is consistent with expectations due to lower bandwidth as the angle decreases. 

Though we are limited to a higher density range in AC measurements of d$\mu$/d$n$, we can directly measure $\mu(n)$ on DC timescales. While this measurement has a higher noise floor, it allows us to measure this sample down to lower densities \cite{foutty_tunable_2023}. In Fig. \ref{fig:SampleB}e, we show an example of $\mu(n)$ measured at $B=0$ T and $T = 1.6$ K at a location with twist angle $\theta = 0.97^\circ$. Blue arrows indicate thermodynamic gaps [i.e., steps in $\mu(n)$] at $v = -1,-2,-3$, and $-4$. By analyzing similar data at a number of independent spatial positions, we can measure thermodynamic gaps at a variety of twist angles and integers, which are presented in Table \ref{tab:twistangles}. We estimate the gaps measured via DC as having uncertainty of approximately $\pm 2$ meV, an order of magnitude worse than our AC measurements. None of the integer gaps measured via either technique disperse in a magnetic field across the twist angle range studied in Sample B.

\newpage
\onecolumngrid
\section{Supplementary Figures}

 \begin{figure*}[h]
    \renewcommand{\thefigure}{S\arabic{figure}}
    \centering
    \includegraphics[scale=1.0]{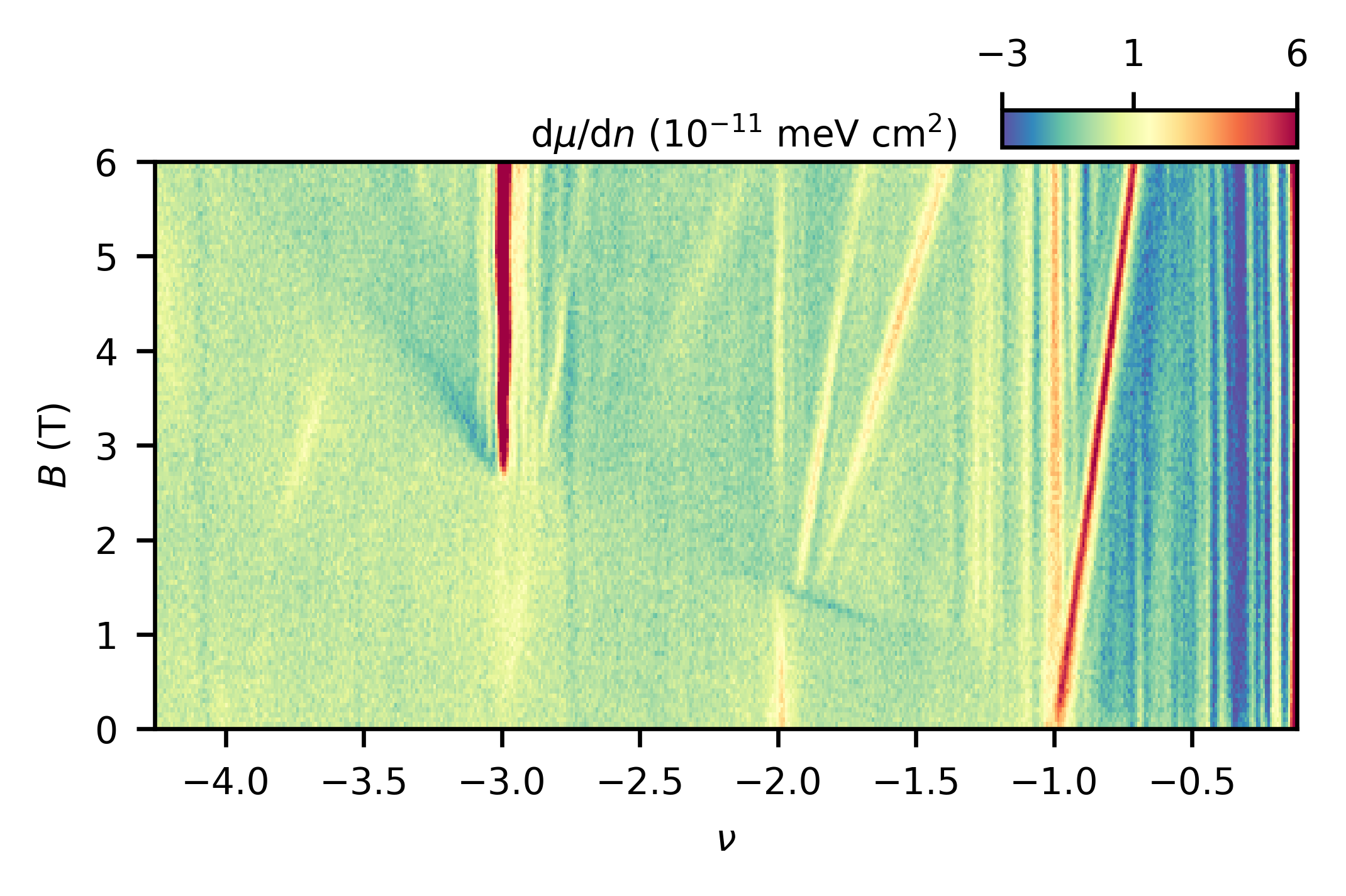}
    \caption{\textbf{d$\mu$/d$n$ at $T = 1.6$ K.} d$\mu$/d$n$ as a function of $\nu$ and $B$ at temperature $T = 1.6$ K. All other parameters (including location in the sample) are identical to the data shown Fig. 2a.}
    \label{fig:1.6K}
\end{figure*}

 \begin{figure*}[h]
    \renewcommand{\thefigure}{S\arabic{figure}}
    \centering
    \includegraphics[scale=1.0]{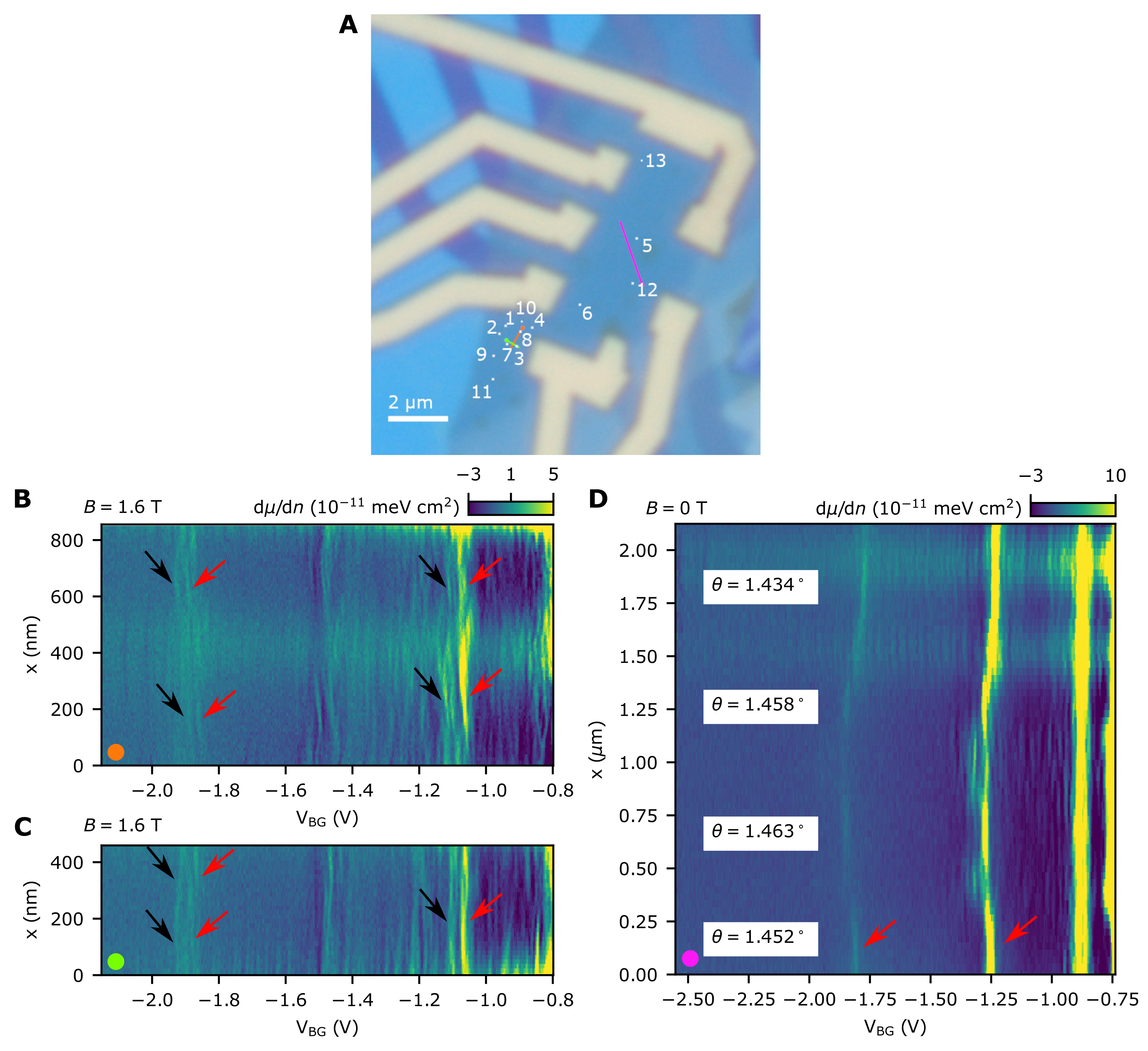}
    \caption{\textbf{Device micrograph and spatial characterization.} \textbf{(A)} Optical micrograph of the device, with individual points marking measurements. Points 1-6 mark the locations of the measurements presented in Fig. 3a in the main text, from the top to bottom. Point 3 is the location where the data in Figs. 1-2 is taken. Points 7-8 mark the locations of the measurements in Supplementary Fig. \ref{fig:AltSpots}. Point 9 marks the location of Fig. 4c-f in the main text and Supplementary Fig. \ref{fig:1.2deg}. Points 10-13 are clean locations twisted to $\theta = 1.25^\circ$, $1.35^\circ$ $1.45^\circ$, and $1.60^\circ$ respectively that are included in Figs. 3b-c and 4g and shown in Supplementary Fig. \ref{fig:Othertwists}. Orange, green, and pink lines show the locations of data plotted in (B-D), where the dot indicates $x=0$ for each line.  \textbf{(B,C)} d$\mu$/d$n$ as a function of back gate voltage $V_{BG}$ and spatial position $x$ along two roughly orthogonal lines in real space (shown in (A)) spanning a uniform region with $\theta \approx 1.23^\circ$, at a constant magnetic field of $B = 1.6$ T. Spatial variability in the location of features is mostly due to potential disorder in the sample rather than twist angle disorder. Features related to $\nu = -1,-2,$ and $-3$ are clustered around $V_{BG} = -1.1,-1.5$, and $-1.9$ V, respectively. The dominance of the $C= +1$ state (relative to $C=0$ at $\nu = -1$) is consistent spatially, and coexistence of both is also visible near $\nu = -3$. $C = +1$ gaps are highlighted by red arrows, while $C = 0$ gaps are highlighted by black arrows. While these spatial line cuts respectively span approximately 500 nm and 800 nm, spot checks in other locations indicate qualitatively uniform behavior throughout a 600 nm $\times$ 1 $\mu$m region.  \textbf{(D)}, d$\mu$/d$n$ at $T = 1.6$ K and $B = 0$ T as a function of position along a line through a higher-twist region of the device.  The red arrows indicate the $\nu = -1$ and $\nu = -2$ gaps. These shift in gate voltage as the twist angle changes. The twist angle derived from the density difference between the pair of integer states for prominent domains is annotated on the left. We observe domains with a typical size of about 300 nm, over which the system is highly homogeneous as sensed by our $\sim 100$ nm probe.} 
    \label{fig:DeviceCharacterization}
\end{figure*}

 \begin{figure*}[h]
    \renewcommand{\thefigure}{S\arabic{figure}}
    \centering
    \includegraphics[scale=1.0]{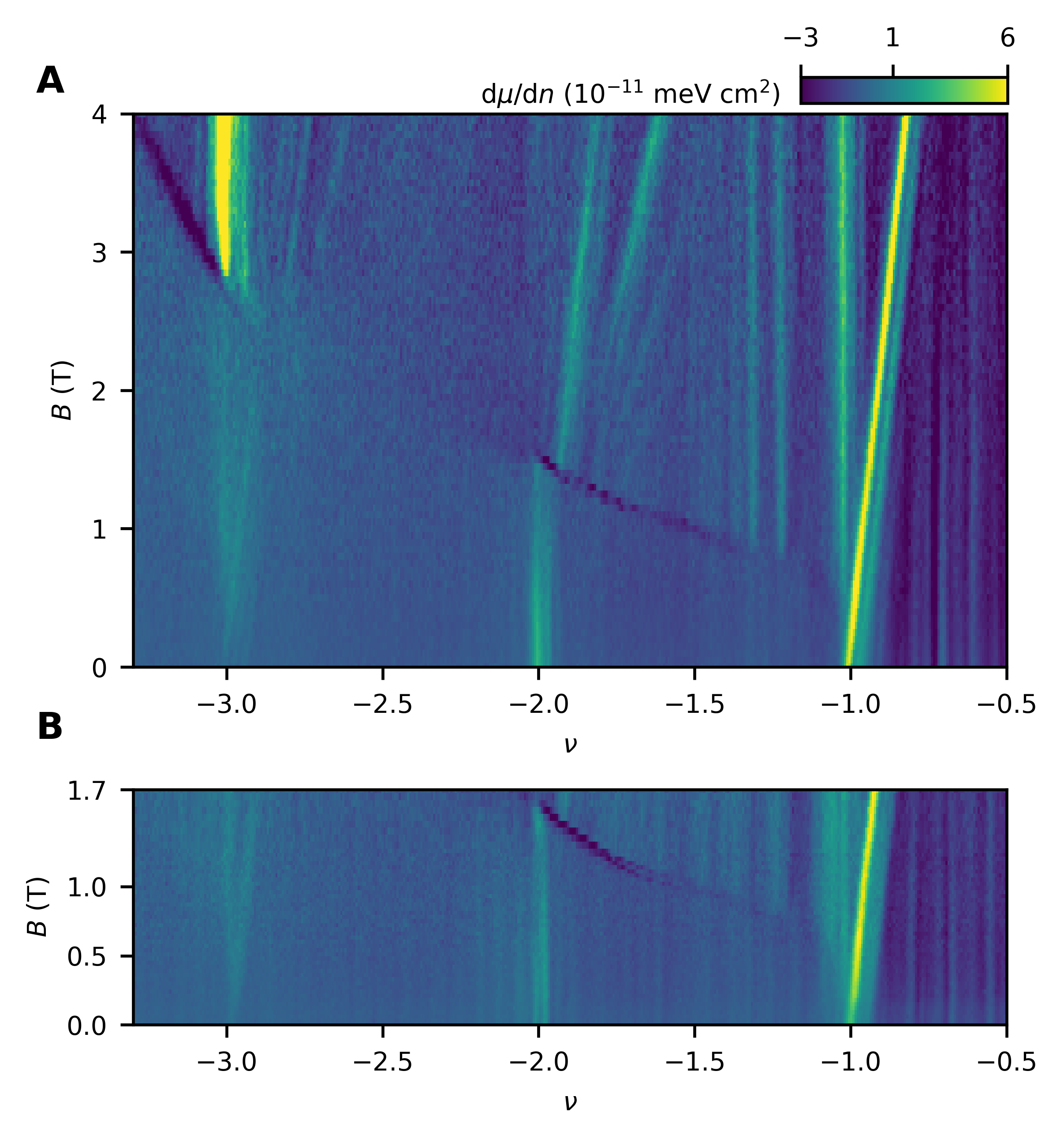}
    \caption{\textbf{Topological gaps at other independent locations.} \textbf{(A,B)} d$\mu$/d$n$ as a function of $\nu$ and $B$ in two independent locations in the sample. These locations are separated from the data shown in Fig. 1 in the main text and from each other by $>500$ nm. The twist angle is roughly the same (within $\pm 0.005^\circ$) across the three locations.} 
    \label{fig:AltSpots}
\end{figure*}

 \begin{figure*}[h]
    \renewcommand{\thefigure}{S\arabic{figure}}
    \centering
    \includegraphics[scale=1.0]{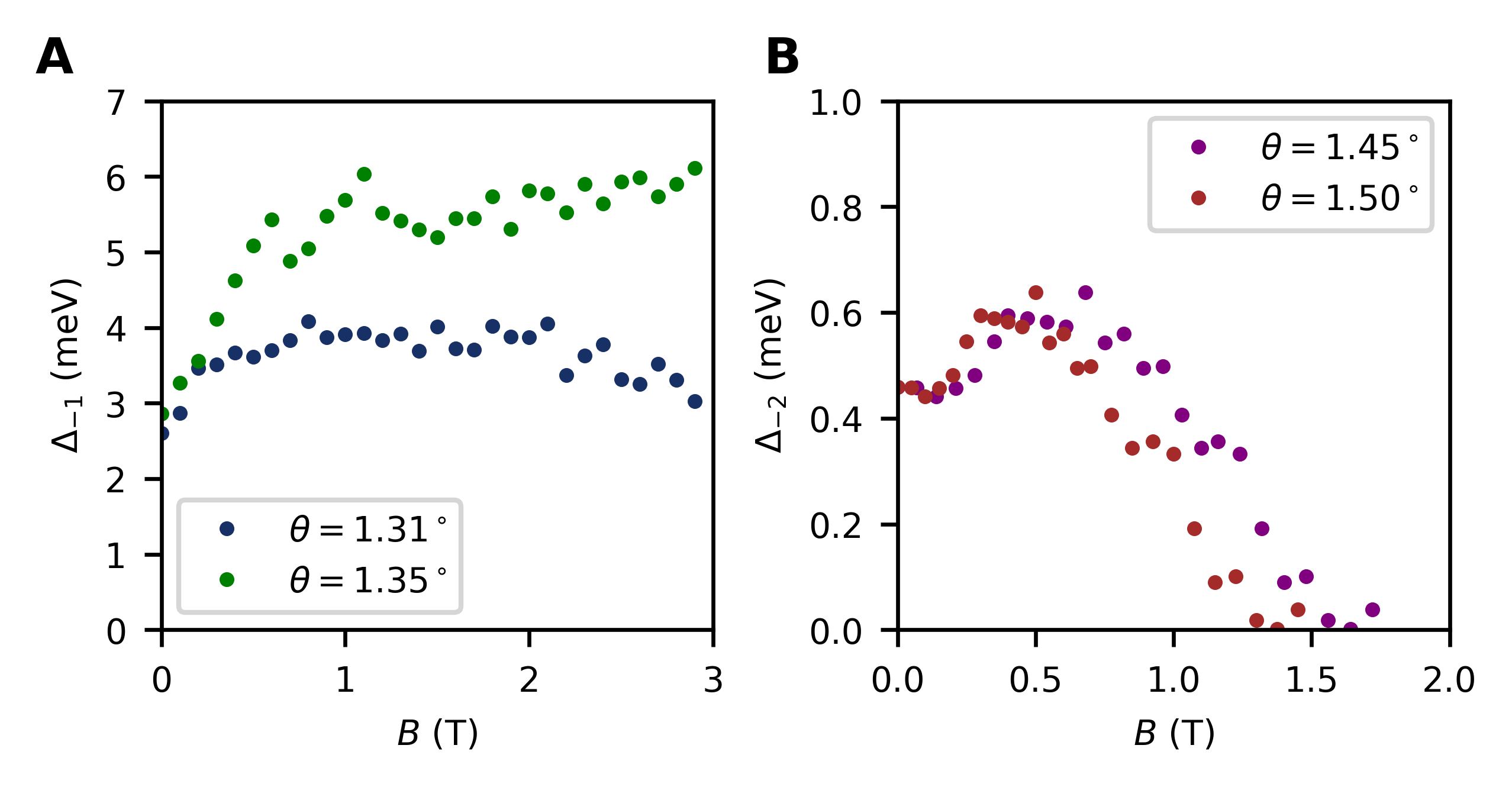}
    \caption{\textbf{Field dependence of thermodynamic gaps at non-topological twist angles.} \textbf{(A)} Gap sizes of the $(C,s) = (0,-1)$ gaps at $\theta = 1.31^\circ$ and $\theta = 1.35^\circ$. \textbf{(B)} Gap sizes of the $(C,s) = (0,-2)$ gaps at $\theta = 1.45^\circ$ and $\theta = 1.50^\circ$. Both gaps show non-monotonic gap size dependence on magnetic field, and there is no evidence for topological bands at these twist angles.} 
    \label{fig:GapSizesOthertwists}
\end{figure*}

 \begin{figure*}[h]
    \renewcommand{\thefigure}{S\arabic{figure}}
    \centering
    \includegraphics[scale=1.0]{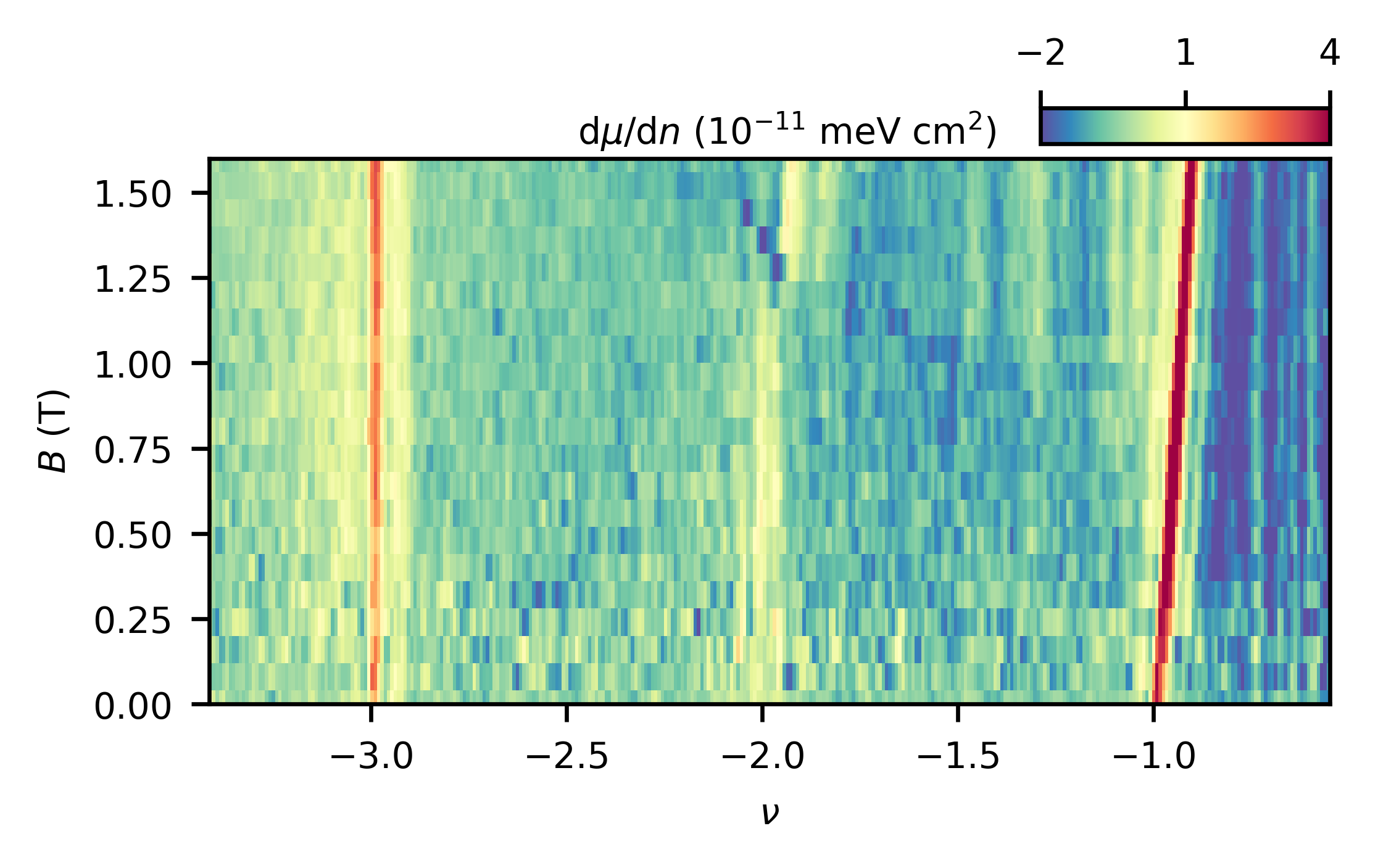}
    \caption{\textbf{Low-field measurement at $\theta = 1.20^\circ$}. d$\mu$/d$n$ as a function of $\nu$ and $B$ at low magnetic fields, in the same spot as the measurements in Fig. 4c-f in the main text, but with no tip doping. This is the only twist angle where at $B=0$, we resolve a clear $C=+1$ gap at $\nu = -1$ but a $C = 0$ state at $\nu = -3$ without changing the displacement field.}
    \label{fig:1.2deg}
\end{figure*}

 \begin{figure*}[h]
    \renewcommand{\thefigure}{S\arabic{figure}}
    \centering
    \includegraphics[scale=1.0]{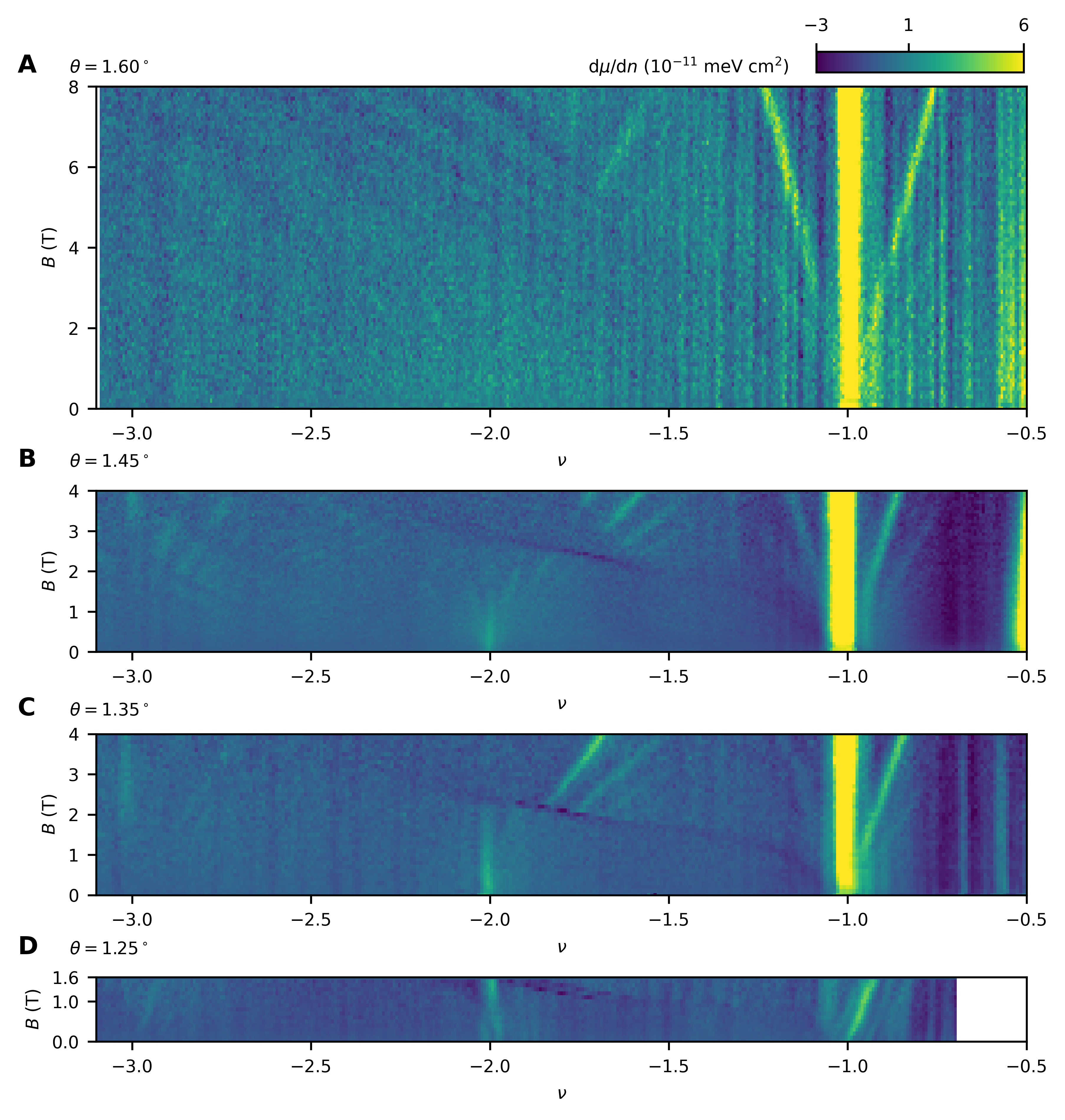}
    \caption{\textbf{Inverse compressibility data at other twist angles.} \textbf{(A-D)}, d$\mu$/d$n$ as a function of $\nu$ and $B$ at four locations at different twist angles from those presented in Fig. 3a in the main text, $\theta = 1.60^\circ$ (A), $\theta = 1.45^\circ$ (B), $\theta = 1.35^\circ$ (C), and $\theta = 1.25^\circ$ (D).} 
    \label{fig:Othertwists}
\end{figure*}

 \begin{figure*}[h]
    \renewcommand{\thefigure}{S\arabic{figure}}
    \centering
    \includegraphics[scale=1.0]{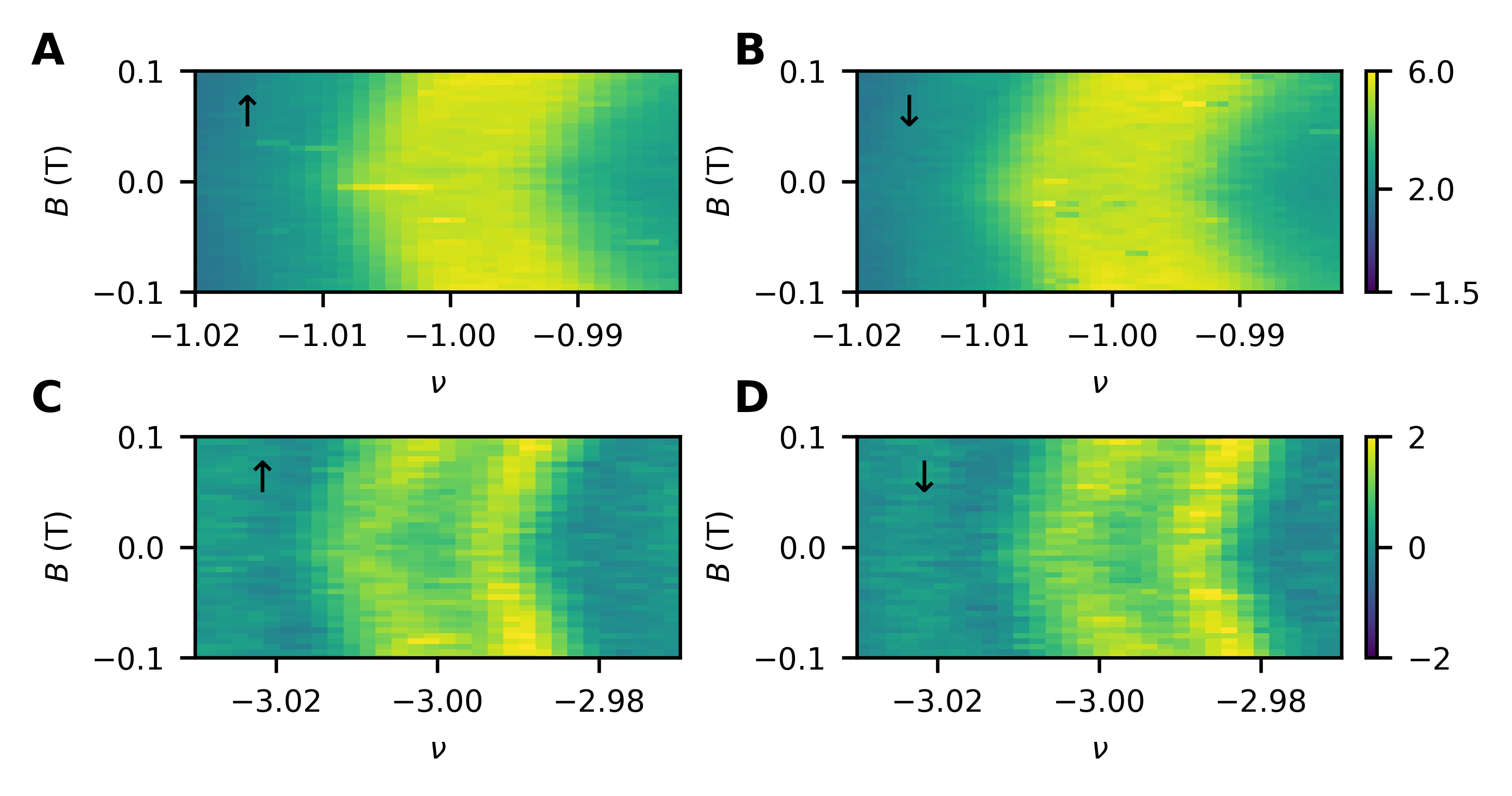}
    \caption{\textbf{Low-field behavior of the gaps from moir\'e filling factors $\nu = -1$ and $\nu = -3$.} \textbf{(A-B)}, Inverse electronic compressibility d$\mu$/d$n$ of the quantum anomalous Hall state emanating from $\nu = -1$ in the same location as Fig. 1-2 of the main text. Arrows indicate whether the magnetic field $B$ is swept up (A) or down (B). Both panels share a colorbar, at right. \textbf{(C-D)}, d$\mu$/d$n$ near the $\nu = -3$ state at the same location. Arrows indicate field sweep direction: up (C) and down (D), and both panels share a colorbar, at right.} 
    \label{fig:Hysteresis}
\end{figure*}

\begin{figure}[t!]
    \renewcommand{\thefigure}{S\arabic{figure}}
    \centering
    \includegraphics[width =\textwidth]{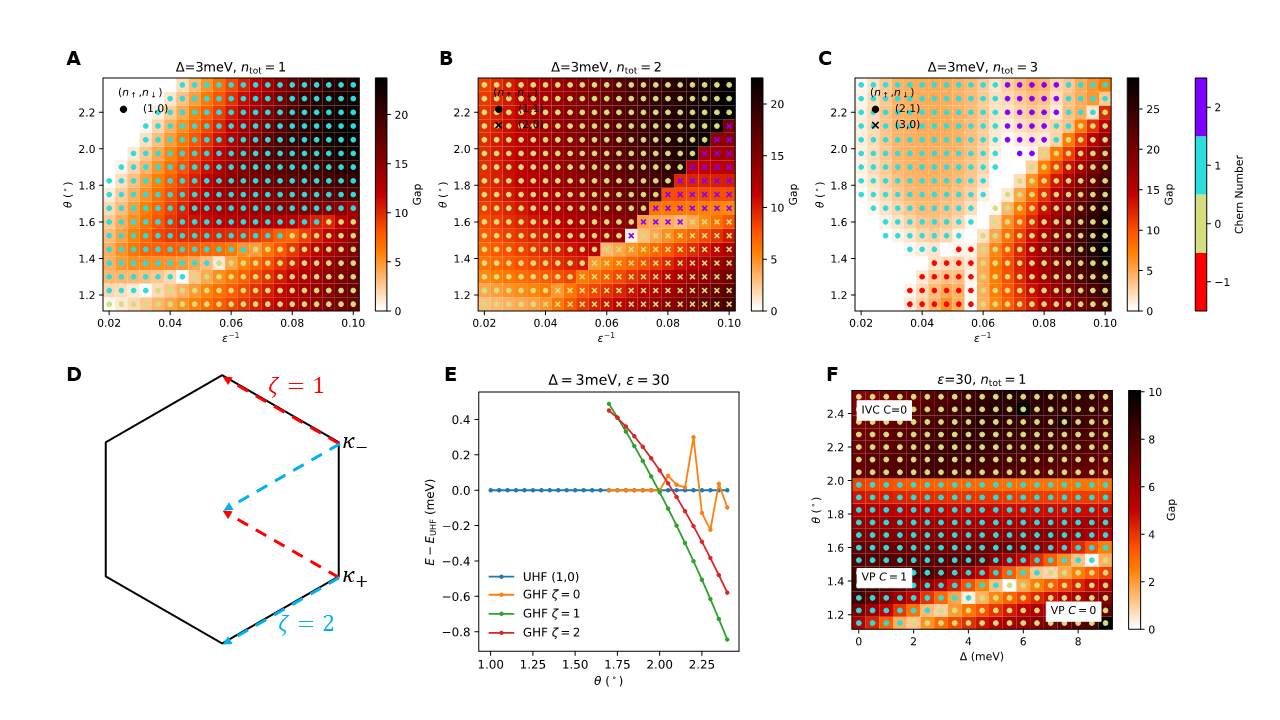}
    \caption{\textbf{Self-consistent Hartree-Fock analysis}. 
    \textbf{a-c}, Self-consistent Hartree-Fock (HF) phase diagrams at a fixed layer potential $\Delta=3$ meV for filling factors as a function of twist angle $\theta$ and dielectric screening $\epsilon$ for $n_{tot}=1$ (\textbf{a}), $n_{tot}=2$ (\textbf{b}), and $n_{tot}=3$ (\textbf{c}).
    \textbf{d}, The $\kappa_{\pm}$ points in the moir\'e BZ (corresponding to $\zeta=0$, and their shifts for the $\zeta=1,2$ choices.
    \textbf{e}, The energy of ``generalized'' HF (GHF) states with $\zeta=0,1,2$ compared to the $(n_{\uparrow},n_{\downarrow})=(1,0)$ ``unrestricted'' HF (UHF) state as a function of $\theta$ for a fixed $\Delta$ and $\epsilon$.
    \textbf{f}, The $n_{tot}=1$ phase diagram as a function of $\theta$ and $\Delta$, including the intervalley coherent (IVC) state.
    } 
    \label{fig:HFFig}
\end{figure}

\begin{figure}[h!]
    \renewcommand{\thefigure}{S\arabic{figure}}
    \centering
    \includegraphics[width=\textwidth]{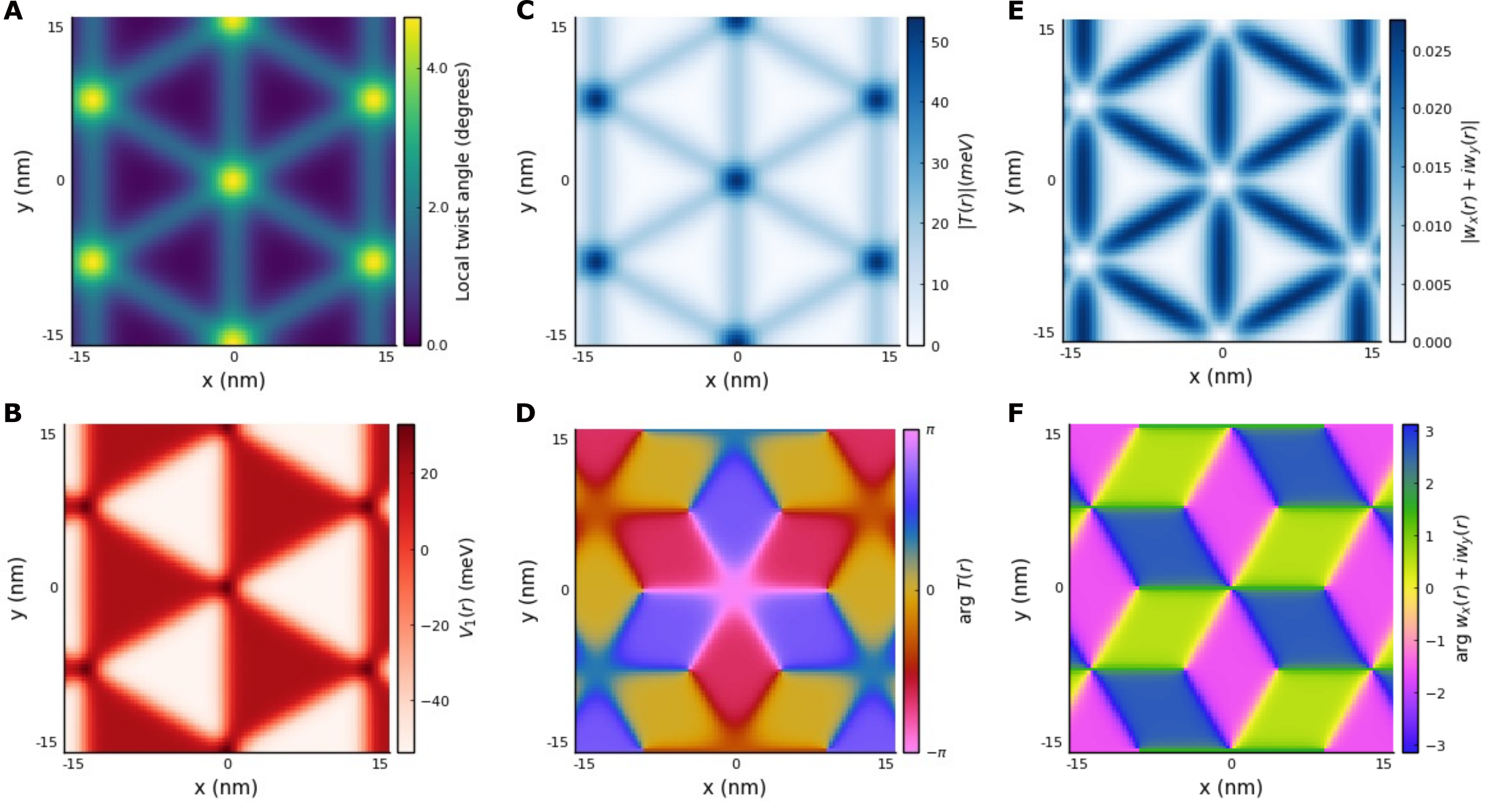}
    \caption{\textbf{Illustration of terms in the relaxed continuum model
    for a heavily relaxed structure with domain wall width $\xi=2$nm.}
    (\textbf{A}), The local twist angle is shown, demonstrating the relaxation towards large domains of zero-twist MX and XM regions.  (\textbf{B-F}), Plots of various quantities for relaxed potential, tunneling, and kinetic terms.
    } 
    \label{fig:RelaxationFig}
\end{figure}

\begin{figure}[h!]
    \renewcommand{\thefigure}{S\arabic{figure}}
    \centering
    \includegraphics{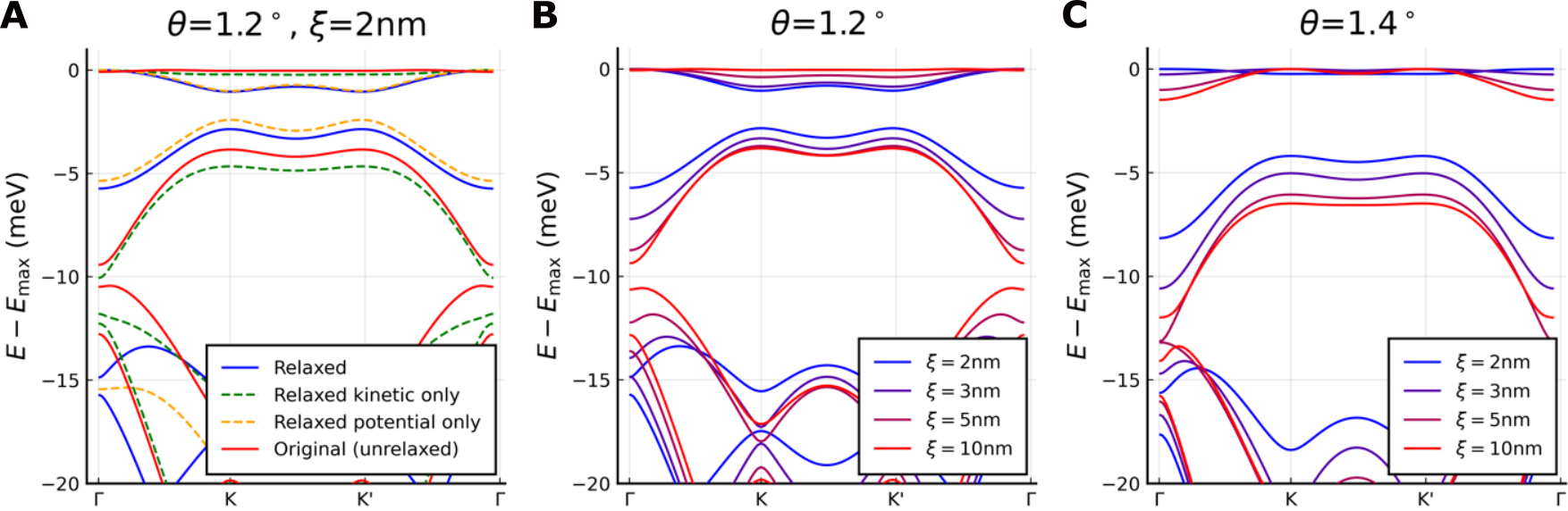}
    \caption{\textbf{Electronic band structure of the relaxed continuum model.}
    (\textbf{A}), The unrelaxed and relaxed band structure for $\xi=2$nm is shown (as well as the band structure including only the potential and tunneling terms, and only the kinetic terms).
    (\textbf{B-C}) The relaxed band structure computed at $\theta=1.2^\circ$ and $\theta=1.4^\circ$ for several choices of $\xi$ from heavily relaxed (2~nm) to almost unrelaxed (10~nm)
    } 
    \label{fig:RelaxationBandsFig}
\end{figure}

\begin{figure}[h!]
    \renewcommand{\thefigure}{S\arabic{figure}}
    \centering
    \includegraphics{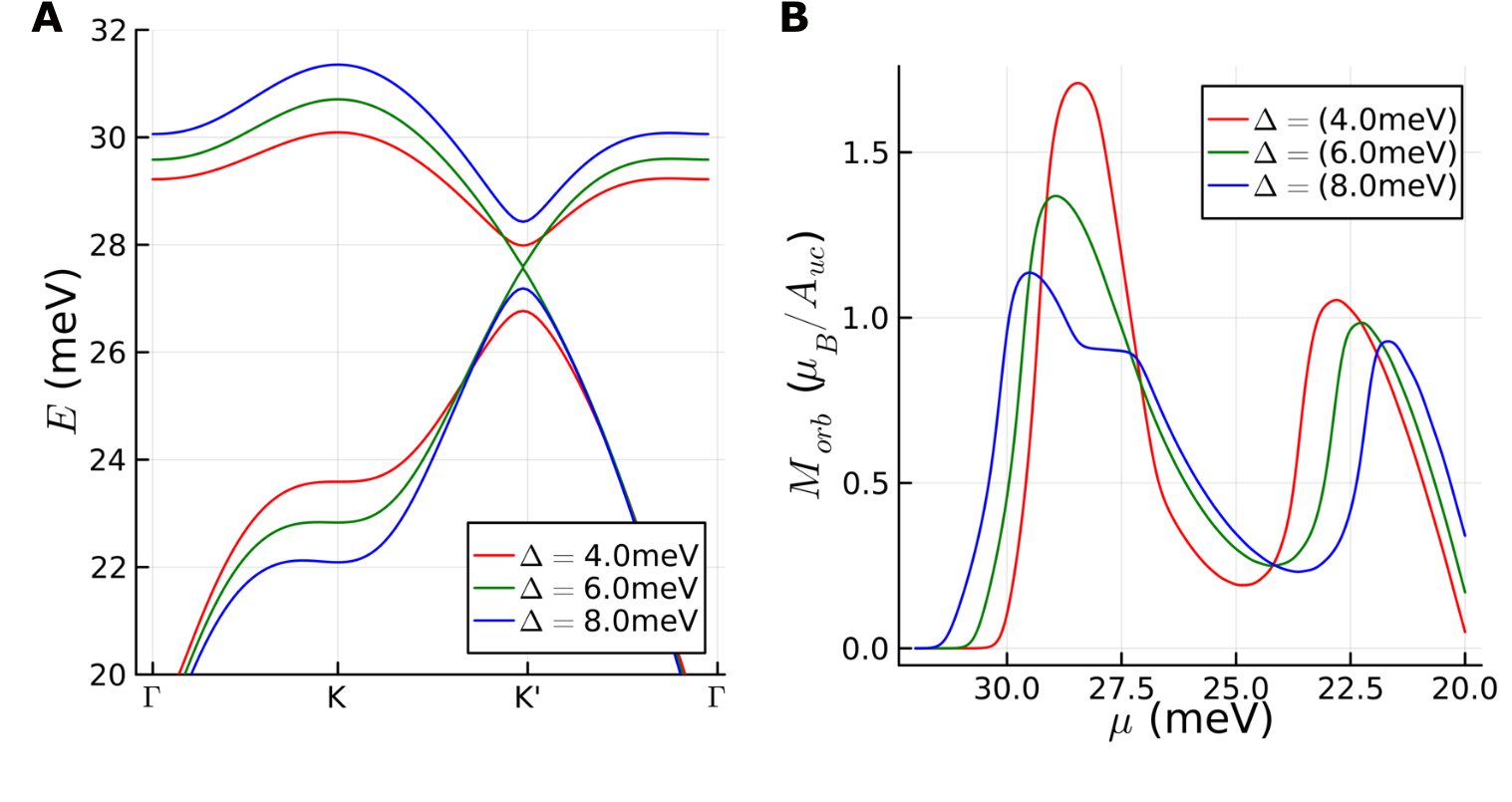}
    \caption{\textbf{Topological band inversion and moir\'e orbital magnetization.}
    (\textbf{A}), The band structure of the continuum model is shown for $\theta=1.2^\circ$ at $\Delta=4,6,8$meV, through which a topological band inversion occurs.
    (\textbf{B}), The orbital magnetization, assuming full valley polarization, is shown for the same choices of $\Delta$ as a function of chemical potential $\mu$.
    } 
    \label{fig:MagnetizationBandsFig}
\end{figure}

 \begin{figure*}[h]
    \renewcommand{\thefigure}{S\arabic{figure}}
    \centering
    \includegraphics[scale=1.0]{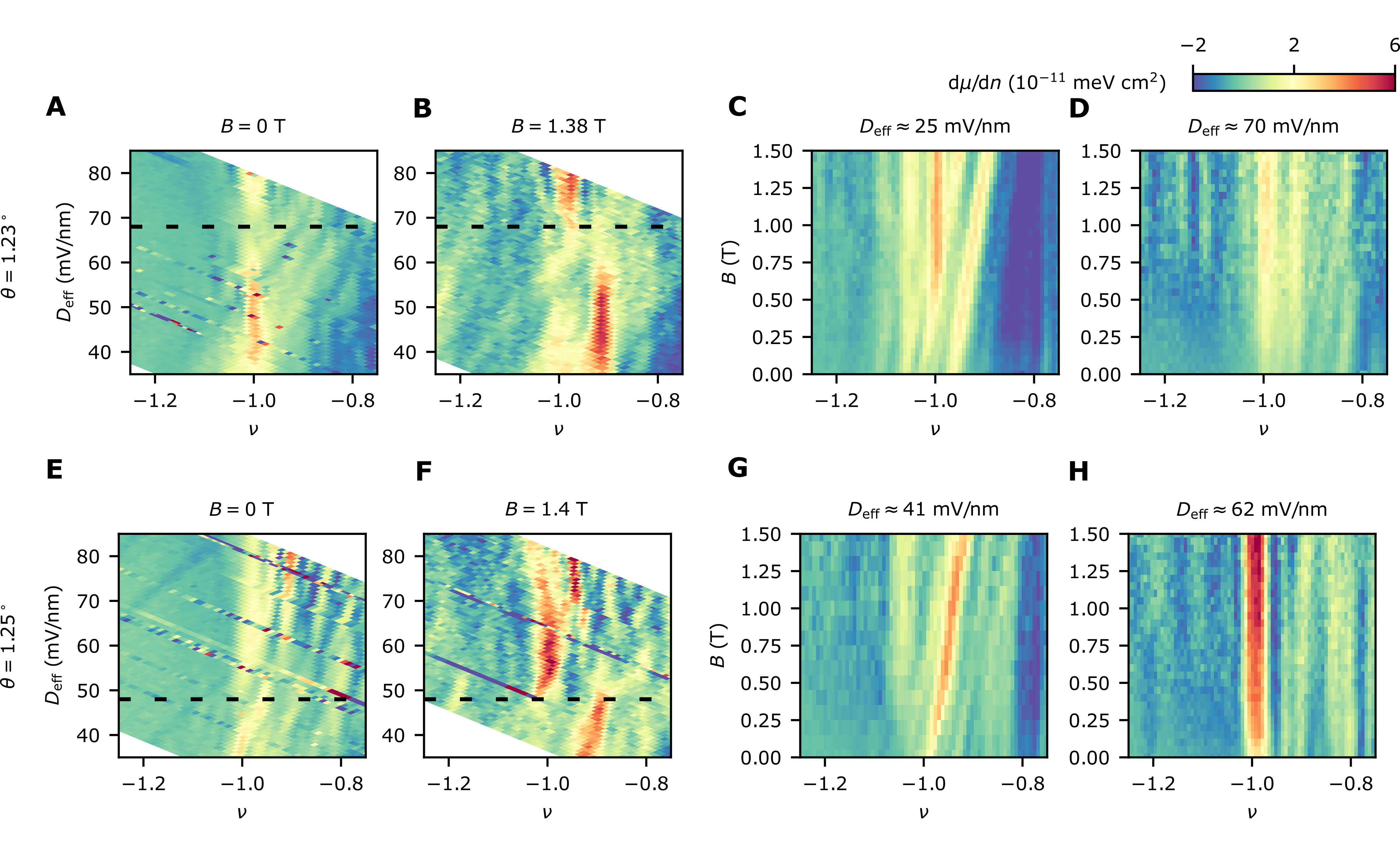}
    \caption{\textbf{Displacement field tuning of $\nu = -1$ gap as a function of twist angle.} \textbf{(A-B)}, d$\mu$/d$n$ in a location with $\theta  = 1.23^\circ$ as a function of $\nu$ and effective displacement field $D_{\rm{eff}}$ at $B = 0$ T (A) and $B = 1.38$ T (B). Black dashed lines indicate $D_{c}\approx 68$ mV/nm for this angle.  \textbf{(C-D)} d$\mu$/d$n$ as a function of $\nu$ and $B$ at $D_{\rm{eff}} \approx 25$ mV/nm (state sloped with $C = +1$ at low magnetic fields, (C)) and $D_{\rm{eff}} \approx 70$ mV/nm (state vertical with $C = 0$ at low magnetic fields, (D)). \textbf{(E-F)}, Similar measurements, performed at a different location with $\theta = 1.25^\circ$. From measurements as a function of $\nu$ and $D_{\rm{eff}}$ at $B = 0$ (E) and $B = 1.4$ T (F), we identify $D_c \approx 48$ mV/nm. \textbf{(G-H)}, From measurements as a function of $\nu$ and $B$ at approximately fixed $D_{\rm{eff}}$, we identify topological behavior at low displacement fields ($D_{\rm{eff}} \approx 41$ mV/nm, (G)) and trivial behavior at higher displacement fields ($D_{\rm{eff}} \approx 62$ mV/nm, (H)).}
    \label{fig:1.23_1.25deg}
\end{figure*}

 \begin{figure*}[h]
    \renewcommand{\thefigure}{S\arabic{figure}}
    \centering
    \includegraphics[scale=1.0]{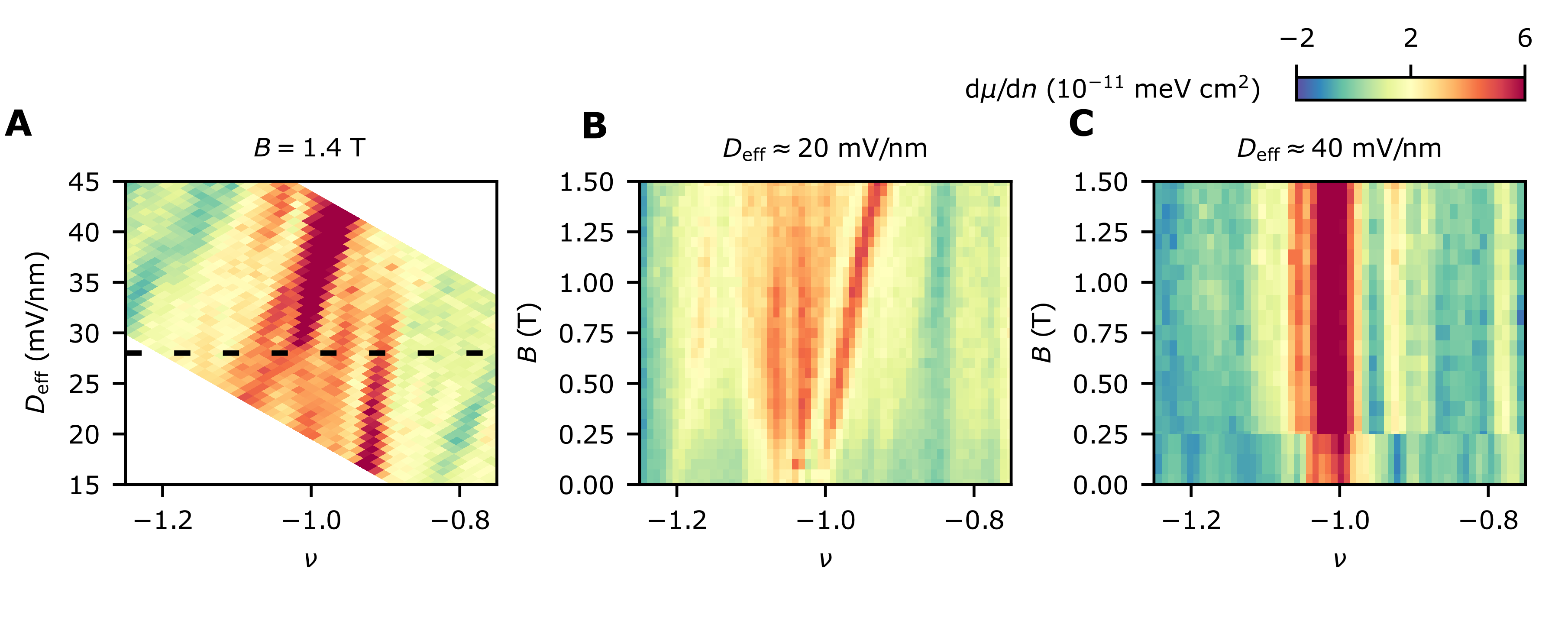}
    \caption{\textbf{Displacement field tuning of $\nu = -1$ at $\theta = 1.19^\circ$.} \textbf{(A)}, d$\mu$/d$n$ as a function of $D_{\rm{eff}}$ and $\nu$ at $B = 1.4$ T. The black dashed line marks $D_{c}$ for this angle, where the dominant state switches from $C = +1$ (below $D_c$) to $C = 0$ (above $D_c$). \textbf{(B)}, d$\mu$/d$n$ as a function of $\nu$ and $B$ up to $B = 1.5$ T at $D_{\rm{eff}} \approx 20$ mV/nm. The most prominent state has slope $C = +1$, and the ``doubled" states to the left, likely coming from inhomogeneous (lower) doping further away from the tip, also have nonzero slope at the lowest magnetic fields. \textbf{(C)}, d$\mu$/d$n$ as a function of $\nu $ and $B$ up to $B = 1.5$ at $D_{\rm{eff}} \approx 40$ mV/nm, at the condition where the tip does not dope the sample.}
    \label{fig:1.185deg}
\end{figure*}

\begin{figure}[h!]
    \renewcommand{\thefigure}{S\arabic{figure}}
    \centering
    \includegraphics[scale =1]{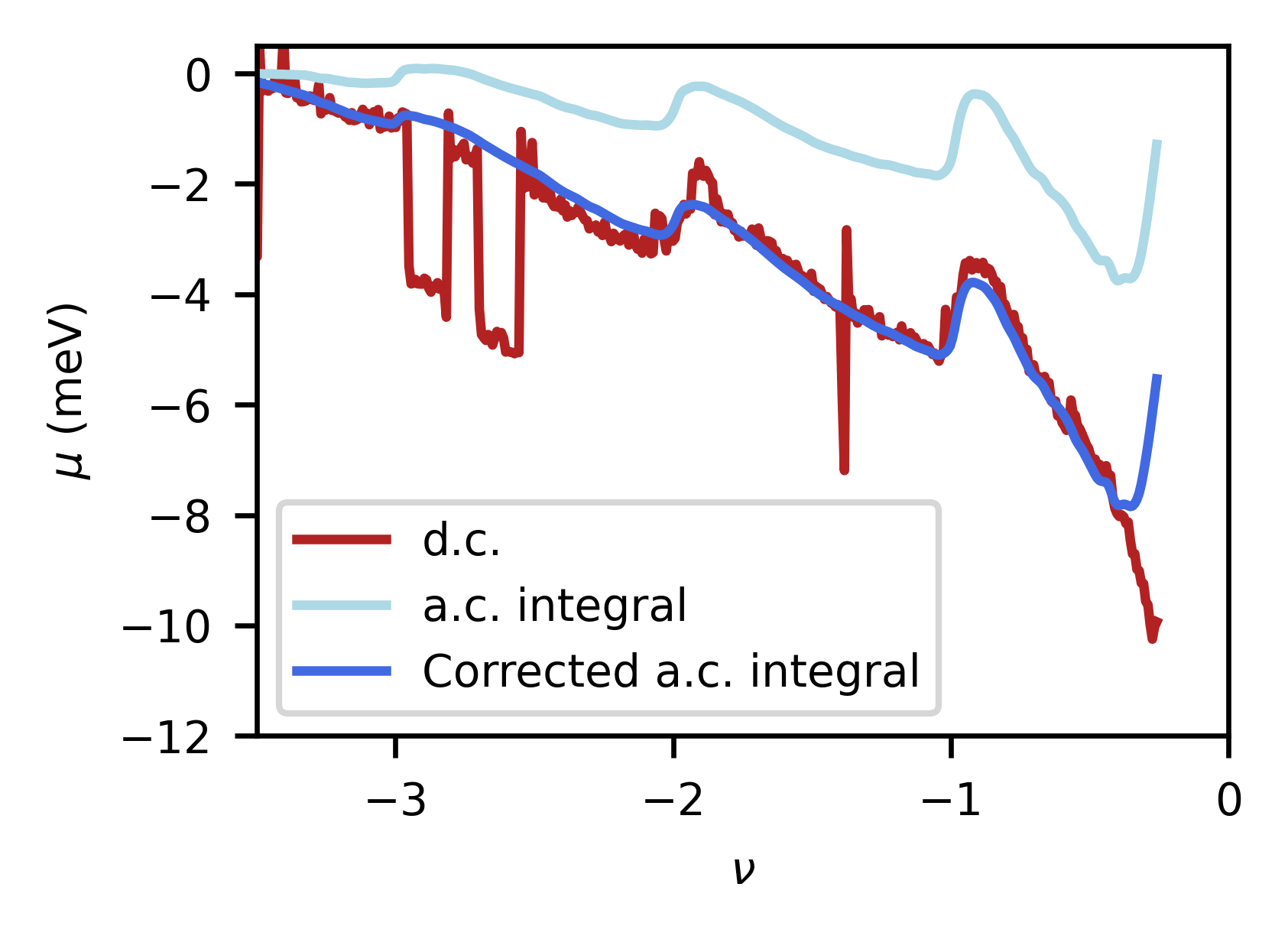}
    \caption{\textbf{$\mu(n)$ Background subtraction}. In red, we present the d.c. measured $\mu(n)$. Discrete jumps (for example around $\nu=-3$) come from telegraphic charge noise in our tip, which does not affect the a.c. measurement. In light blue, we present the integral of a.c. measured d$\mu$/d$n$ without subtraction. The dark blue curve is the integral of d$\mu$/d$n$ with a constant value of $2.7\times 10^{-12}$ meV cm$^{2}$ subtracted before integration, to minimize the least squares difference between the integral and the independently measured d.c. curve.} 
    \label{fig:Background}
\end{figure}

\begin{figure}[h!]
    \renewcommand{\thefigure}{S\arabic{figure}}
    \centering
    \includegraphics[scale =1]{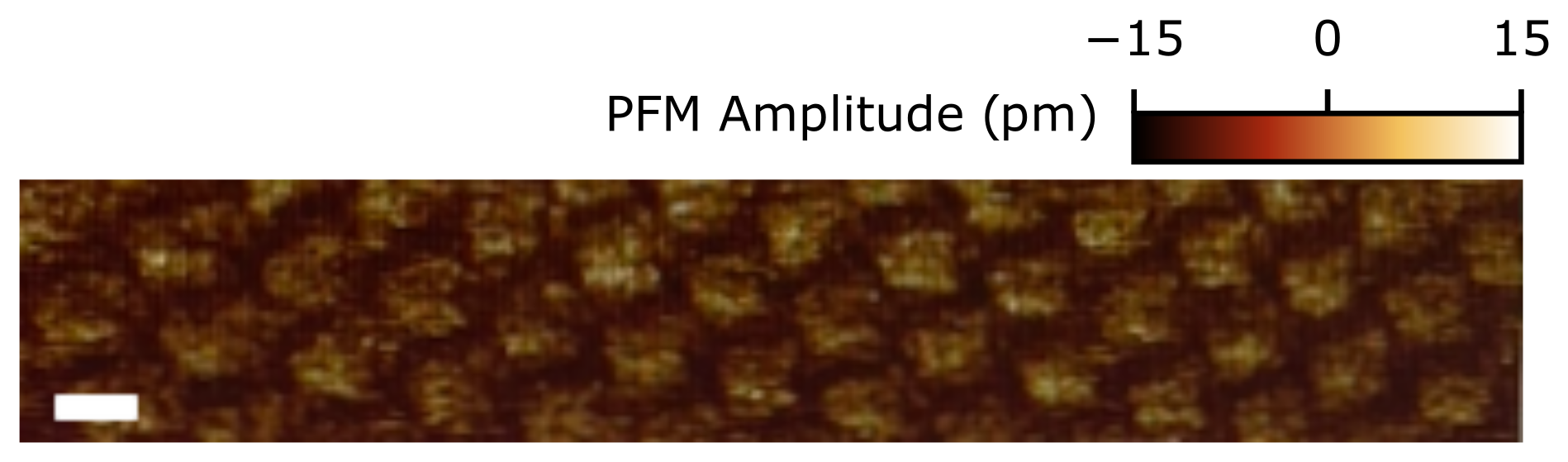}
    \caption{\textbf{PFM Characterization}. Piezoresponse force microscopy (PFM) image taken during the stacking process on the ``open-faced" PC-hBN-tWSe2 stack (Methods), at approximately location 5 in Supplementary Fig. \ref{fig:DeviceCharacterization}. The contrast shows a triangular lattice with an approximate period of $\sim 13$ nm, matching the moire unit cell density that we measure with SET measurements.} 
    \label{fig:PFM}
\end{figure}

\begin{figure}[h!]
    \renewcommand{\thefigure}{S\arabic{figure}}
    \centering
    \includegraphics[scale =1]{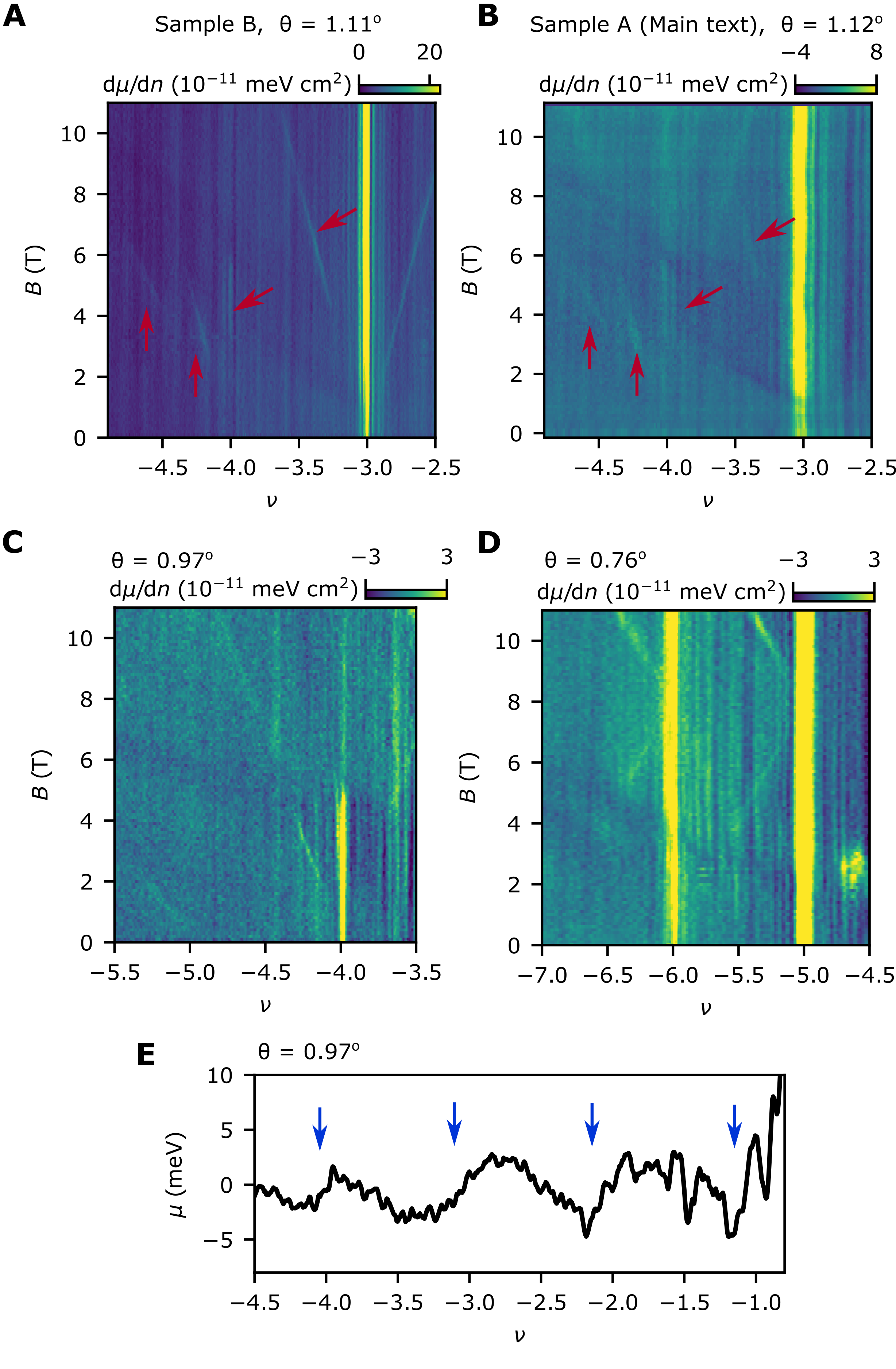}
    \caption{\textbf{Inverse compressibility data from Sample B.}. \textbf{(A-B)} d$\mu$/d$n$ measured as a function of $\nu$ and $B$ at a location with $\theta = 1.11^\circ$ in Sample B (A), alongside data at a similar twist angle ($\theta =1.12^\circ$) in Sample A (B). The red arrows indicate Hofstadter features that are common between the two measurements. \textbf{(C-D)}, d$\mu$/d$n$  measured as a function of $\nu$ and $B$ at far-separated ($>1\ \mu$m apart) locations in Sample B with $\theta = 0.97^\circ$ (C) and $\theta = 0.76^\circ$ (D). \textbf{(E)}, Measurement of $\mu(n)$ as a function of $\nu$ at $B = 0$ T in a location with $\theta = 0.97^\circ$. Data in panels A-E are at $T = 1.6$ K.  }
    \label{fig:SampleB}
\end{figure}

\clearpage
\section{Supplementary Table}

\begin{table}[h]

\renewcommand{\thetable}{S\arabic{table}}

    \centering
    \begin{tabular}{|c|c|c|>{\centering}p{3cm}|c|c|c|}
    \hline
        Twist angle & $\nu = -1$ (meV) & $\nu = -2$ (meV) & $\nu = -3$ (meV) & $\nu = -4$ (meV) & $\nu = -5$ (meV) & $\nu = -6$ (meV)\\ \hline
        $1.11^\circ$ & $8 (\pm 2)$ & $0.4 (\pm 0.2)$ & $3.5 (\pm 0.2)$ & No gap & No gap & No gap \\
        \hline
        $1.06^\circ$ & $8 (\pm 2)$ & $4 (\pm 2)$ & $6 (\pm 0.2)$ & No gap & No gap & No gap \\
       \hline
        $0.97^\circ$ & $9 (\pm 2)$ & $7 (\pm 2)$ & $6 (\pm 0.2)$ & $1 (\pm 0.2)$ & No gap & No gap \\      
       \hline
        $0.76^\circ$ & Not accessible & $11 (\pm 2)$ & $0 (\pm 2)$ \newline (No gap visible in DC, AC measurement not accessible) & $1.4 (\pm 0.2)$ & $1.25 (\pm 0.2)$ & $0.5(\pm 0.2)$ \\        
    \hline

    \end{tabular}
    \caption{Thermodynamic gaps at integer moir\'e filling $\nu$ measured a selection of locations with different twist angles in Sample B. All measurements are conducted at $B = 0$ T and $T = 1.6$ K. Gaps measured using DC measurement of $\mu(n)$ have an estimated uncertainty of 2 meV, while gaps from integrating the AC measurement of d$\mu$/d$n$ have an estimated uncertainty of 0.2 meV. The $v = -1$ gap at the lowest twist angle is not accessible because it occurs at very low densities where even the DC charging of the sample is not reliable.} 

    \label{tab:twistangles}
\end{table}

\clearpage
